\documentclass[12pt]{article}

\catcode`\@=11
\@addtoreset{equation}{section}

\global\arraycolsep=2pt
\usepackage{mathrsfs,amsbsy,amssymb,latexsym,amsfonts,amsmath,cite}
\usepackage{graphicx,color}

\allowdisplaybreaks

\begin{document}

\begin{flushright}
\parbox{4cm}
{KUNS-2509}
\end{flushright}

\vspace*{1.5cm}

\begin{center}
{\Large \bf A new coordinate system for $q$-deformed AdS$_5 \times $S$^5$\\ 
and classical string solutions}
\vspace*{1.5cm}\\
{\large Takashi Kameyama\footnote{E-mail:~kame@gauge.scphys.kyoto-u.ac.jp} 
and Kentaroh Yoshida\footnote{E-mail:~kyoshida@gauge.scphys.kyoto-u.ac.jp}} 
\end{center}

\vspace*{0.25cm}

\begin{center}
{\it Department of Physics, Kyoto University, \\ 
Kyoto 606-8502, Japan} 
\end{center}

\vspace{2cm}

\begin{abstract}
We study a GKP-like classical string solution on a 
$q$-deformed AdS$_5\times$S$^5$ background and 
argue the spacetime structure by using it as a probe. 
The solution cannot stretch beyond the singularity surface 
and this result may suggest that the holographic relation is realized inside 
the singularity surface. This observation leads us to introduce a new coordinate system 
which describes the spacetime only inside the singularity surface.  
With the new coordinate system, we study minimal surfaces 
and derive a deformed Neumann-Rosochatius system with a rigid string ansatz. 
\end{abstract}

\setcounter{footnote}{0}
\setcounter{page}{0}
\thispagestyle{empty}

\newpage

\section{Introduction}
The AdS/CFT correspondence \cite{M} is one of the central subjects in string theory. 
The most well-studied example is the duality between type IIB string theory on AdS$_5\times$S$^5$ 
background and $\mathcal{N}=4$ $SU(N)$ super Yang-Mills theory in four dimensions. 
A remarkable feature of this duality is integrability \cite{review} and many significant 
properties have been revealed based on exact computations. 
Hence this duality is considered to play a role like `harmonic oscillators' 
in studies of fascinating relations between gauge theories and gravitational theories 
(string theories). 

\medskip 

From a point of view of string theory, the integrability is closely related 
to the integrable structure of non-linear sigma models in two dimensions. 
For type IIB superstring theory defined on AdS$_5\times$S$^5$ 
(often called the AdS$_5\times$S$^5$ superstring), 
the classical action is constructed based on a supersymmetric coset \cite{MT}, 
\[
\frac{PSU(2,2|4)}{SO(1,4)\times SO(5)}\,. 
\] 
It enjoys the $\mathbb{Z}_4$-grading and the classical integrability is shown 
by Bena-Polchinski-Roiban \cite{BPR}\footnote{For another argument based on 
the Roiban-Siegel formulation \cite{RS}, see \cite{Hatsuda}.}. 
Possible supercosets, which lead to consistent string backgrounds, 
are classified in \cite{Zarembo-symmetric}. 

\medskip 

It is interesting to consider integrable deformations of the AdS$_5\times$S$^5$ 
superstring. It opens up a new perspective on the integrable structure of 
the gauge/gravity correspondence beyond the usual AdS/CFT case. 
A good way along this direction is to follow the Yang-Baxter sigma model 
description proposed by Klimcik \cite{Klimcik}. Then, an integrable deformation is specified 
by picking up a classical $r$-matrix satisfying the modified classical Yang-Baxter equation 
(mCYBE). The original argument was restricted to principal chiral models, 
but it is now generalized to the symmetric coset case \cite{DMV}\footnote{
This generalization employed earlier arguments 
on $q$-deformations of $su(2)$ and $sl(2)$ studied in 
a series of papers \cite{KYhybrid,KMY-QAA}. 
For other earlier arguments, see \cite{KOY,Sch}.} 
(For the related progress, see \cite{Hollowood}). 
Just after this success, the classical action of a $q$-deformed AdS$_5\times$S$^5$ 
superstring has been constructed \cite{DMV-string}.  

\medskip 

The pioneering work \cite{DMV-string} has developed a new arena. 
The coordinate system was introduced in \cite{ABF}\footnote{We refer to  it as the ABF metric.
}.
So far, 
the metric in the string frame and NS-NS two-form have been obtained. 
However, the dilaton and R-R sector have not been determined yet and 
it is still an important issue to find out the full solution. In addition, 
there exists a singularity surface in the deformed AdS$_5$ part. 
Some generalizations to AdS$_n\times$S$^n$ have been discussed in \cite{HRT}. 
The mirror theory of the $q$-deformed theory is argued in \cite{mirror,mirror2}. 
The fast-moving string limit offers a possible resolution of the singularity \cite{KY-LL}.
Giant magnon solutions \cite{HM} are constructed in \cite{mirror,Kluson,Ahn}. 
Some classical string solutions have been discussed in \cite{Banerjee,talk}. 
Deformed Neumann models are also obtained in \cite{AM-NR}.

\medskip 

As a generalization of \cite{DMV-string}, one may consider integrable deformations with 
classical $r$-matrices satisfying the classical Yang-Baxter 
equation (CYBE), rather than mCYBE. Along this line, the classical action of the deformed 
AdS$_5\times$S$^5$ superstring was constructed \cite{KMY-JordanianAdSxS}.  
Some examples of classical $r$-matrices \cite{KMY-JordanianIIB,MY1,MY2,CMY-T11} 
have been found for well-known type IIB supergravity solutions 
including the Lunin-Maldacena-Frolov solutions \cite{LM,Frolov} 
and gravity duals for non-commutative gauge theories \cite{HI,MR,DHH}.  

\medskip 

In this paper, we will focus upon the $q$-deformed AdS$_5\times$S$^5$ superstring 
\cite{DMV-string}. An interesting issue is to reveal the nature of the singularity surface. 
For this purpose, we consider a GKP-like classical string solution\footnote{
The solution in AdS$_5$ was originally constructed by 
Gubser-Klebanov-Polyakov (GKP) \cite{GKP}.} as a probe.
Then it is shown that the classical solution cannot be stretched beyond the singularity surface.
This observation may indicate that the holographic relation could be argued 
only inside the singularity surface. Therefore, we propose a new coordinate system 
which describes the spacetime inside the singularity surface, 
rather than taking a truncation limit (i.e., without any approximations). 
Moving to the new coordinate system, we study classical string surfaces 
and derive a deformed Neumann-Rosochatius (NR) system with a rigid string ansatz. 

\medskip 

An adavantage to adopt the new coordinate system is that the unphysical region,  
where the sign of time component of the metric is flipped, 
is not contained.
In the unphysical region, time evolution of strings is not well-defined and hence 
the motion of classical solutions is not described in a standard manner. 
Although the metric is invariant under the time translation even in the unphysical region, 
one cannot define the conservation laws of global charges.  
It is because the Cauchy surfaces are not globally defined and the integral of the charges would be indefinite.
As a matter of course, there are many choices of the coordinate system which describe the spacetime enclosed 
by the singularity surface. Though our choice is just an example, 
it would be useful because it reproduces the usual global coordinates in the undeformed limit.

\medskip

This paper is organized as follows. 
Section 2 considers a GKP-like string solution on the $q$-deformed AdS$_5\times$S$^5$ background.
It is shown that the classical solution cannot stretch beyond the singularity surface.
In section 3, we introduce a new coordinate system and then study minimal surfaces. 
The first example is a static one and its shape is a deformed AdS$_2$\,. 
In the undeformed limit, the solution is reduced to a familiar one \cite{Wilson}. 
A possible generalization is to add an angular momentum.  
This can also be regarded as a deformation of the solution \cite{drukker}\,.
In section 4, we suppose a rigid string ansatz for the deformed S$^5$ part 
and reduce the system to a one-dimensional deformed NR system. 
The resulting system can be rewritten into the canonical form with 
interaction potentials of trigonometric and hyperbolic types. 
Section 5 is devoted to conclusion and discussion. 

\medskip 

Appendix A explains trajectories of massless and massive particles 
in the ABF coordinates and the new coordinates.
In Appendix B,  giant magnons are reconsidered with the new coordinate system.
The dispersion relation is the same as the one with the original coordinates \cite{mirror,Kluson,Ahn}.
In Appendix C, we derive a deformed NR system 
from the deformed AdS$_5$ part.
The resulting system can be rewritten into the canonical form 
with interaction potentials of hyperbolic type.

\section{The $q$-deformed AdS$_5\times$S$^5$}

We first introduce the bosonic part of the $q$-deformed AdS$_5\times$S$^5$ 
superstring action constructed in \cite{DMV-string,ABF}. 
Then a GKP-like string solution is considered on the deformed background. 
In particular, we study two limits of the solution, 1) a short-string limit and 2) a long-string limit.  
In each limit, the classical energy is expressed as an expansion in terms of an angular momentum. 

\subsection{The bosonic part of the AdS$_5\times$S$^5$ superstring action}

In the following, we are concerned with the bosonic part of the $q$-deformed 
AdS$_5\times$S$^5$ superstring action \cite{DMV-string}, 
where the metric (in the string frame) 
and NS-NS two-form have been determined in \cite{ABF}. 
The complete supergravity solution has not been found yet, but the metric and 
NS-NS two-form are sufficient for our present analysis. 

\medskip

The bosonic action (in the conformal gauge) is composed of the metric part $S_G$ 
and the Wess-Zumino (WZ) term $S_{\rm WZ}$ like 
\begin{eqnarray}	
S &=& S_G + S_{\rm WZ}\,. \nonumber 
\end{eqnarray} 
Here $S_{\rm WZ}$ describes the coupling of string to an NS-NS two-form. 

\medskip 

The metric part $S_G$ can be divided into the AdS part and the internal sphere part like 
\begin{eqnarray}	
S_G &=& \int\!d\tau d\sigma\, \left[\,
\mathcal{L}^G_{\rm AdS} + \mathcal{L}^G_{\rm S}\,
\right] \nonumber \\
& = &-\frac{1}{4\pi \alpha'}\int\!\!d\tau d\sigma\,\eta_{\mu\nu}
\left[G^{MN}_{\rm AdS}\partial^{\mu}X_M \partial^{\nu}X_N +
G^{PQ}_{\rm S}\partial^{\mu}Y_P \partial^{\nu}Y_Q \right]\,, \nonumber 
\end{eqnarray}
where the string world-sheet coordinates are $\sigma^{\mu}=(\tau,\sigma)$ with $\eta_{\mu\nu}=(-1,+1)$\,. 
The metric for the AdS part and the sphere part are given by, respectively,  
\begin{eqnarray}
ds^2_{\textrm{AdS}_5} &=& R^2(1+C^2)^{\frac{1}{2}}
\Bigl[ -\dfrac{\cosh^2\rho\,dt^2}{1-C^2\sinh^2\rho}
+\dfrac{d\rho^2}{1-C^2\sinh^2\rho}
+\dfrac{\sinh^2\rho\,d\zeta^2}{1+C^2\sinh^4\rho\sin^2\zeta}\nonumber\\ 
&&\hspace{2.0cm}+\dfrac{\sinh^2\rho\cos^2\zeta\,d\psi_1^2}{1+C^2\sinh^4\rho\sin^2\zeta}
+\sinh^2\rho\sin^2\zeta\,d\psi_2^2\Bigr]\,,\label{ads5}\\\nonumber\\
ds^2_{\textrm{S}^5} &=& R^2(1+C^2)^{\frac{1}{2}}
\Bigl[  \dfrac{\cos^2\theta\,d\phi^2}{1+C^2 \sin^2\theta}
+\dfrac{d\theta^2}{1+C^2\sin^2\theta}
+\dfrac{\sin^2\theta\,d\xi^2}{1+C^2\sin^4\theta\sin^2\xi}\nonumber\\ 
&&\hspace{2.0cm}+\dfrac{\sin^2\theta\cos^2\xi\,d\phi_1^2}{1+C^2\sin^4\theta\sin^2\xi}
+\sin^2\theta\sin^2\xi\,d\phi_2^2\Bigr]\,.
\label{s5}
\end{eqnarray}
Here the deformed AdS$_5$ is parameterized by\footnote{
For our convenience, we have slightly changed the notation of \cite{ABF}. 
The dictionary is as follows: 
\[
	\rho_{\rm ABF} = \sinh\rho_{\rm us}\,,\qquad r_{\rm ABF} = \sin\theta_{\rm us}\,.
\]
There is no essential difference between the two notations. 
} 
the coordinates $(t\,, \psi_1\,, \psi_2\,,\zeta\,, \rho)$\,. 
The deformed S$^5$ is done by $(\phi\,, \phi_1\,, \phi_2 \,, \xi\,, \theta)$\,. 
The deformation is measured by a real parameter $C \in [0,\infty)$\footnote{
One can fix the range of $C$ as $C \geq 0$ without loss of generality. Note that 
the $q$-deformed metric given by (\ref{ads5}) and (\ref{s5}) depends on $C^2$ only  
and the coupling to the NS-NS two-form is linear in $C$\,. Hence, by fixing the signature of 
the string charge, the range $C \geq 0$ is realized in general.
}.
When $C=0$\,, the geometry is reduced to the undeformed AdS$_5\times$S$^5$ 
with the curvature radius $R$\,. 
The range of the coordinates is the same as in the case with $C=0$\,. 

\medskip 

The WZ term is also divided into two parts:    
\begin{eqnarray}
	\mathcal{L}_{\textrm{AdS}}^{WZ} &=& \frac{T_{(C)}}{2}\,C\,
\epsilon^{\mu\nu}\dfrac{\sinh^4\rho\sin2\zeta}{1+C^2\sinh^4\rho\sin^2\zeta}\,
\partial_\mu\psi_1\partial_\nu\zeta\,, 
\label{wz_ads} \\  
	\mathcal{L}_{\textrm{S}}^{WZ} &=& -\frac{T_{(C)}}{2}\,C\,
\epsilon^{\mu\nu}\dfrac{\sin^4\theta\sin2\xi}{1+C^2\sin^4\theta\sin^2\xi}\,
\partial_\mu\phi_1\partial_\nu\xi\,.
\label{wz_s}
\end{eqnarray}
Here $\epsilon^{\mu\nu}$ is the totally anti-symmetric tensor on the world-sheet 
and it is normalized as $\epsilon^{01}=+1$\,. 
A $C$-dependent string tension $T_{(C)}$ is defined as 
\begin{eqnarray}
	T_{(C)}\equiv T(1+C^2)^{\frac{1}{2}}\,, \qquad 
T\equiv \dfrac{R^{2}}{2\pi\alpha'}\,, \qquad 
\sqrt{\lambda}\equiv\dfrac{R^{2}}{\alpha'}\,.
\end{eqnarray}
Each component of the WZ term is proportional to $C$\,, 
hence it vanishes when $C=0$\,.

\subsection{A GKP-like string solution}

Let us consider a GKP-like string solution \cite{GKP} 
on the $q$-deformed AdS$_5\times$S$^5$ background. 
The solution describes a folded, rotating string sitting 
at the origin of the deformed AdS$_5$\,. 
As a remarkable feature, it cannot stretch beyond the singularity surface 
by changing the value of an angular momentum. 
This part has some overlap with \cite{talk}. 

\subsubsection*{A rotating string ansatz}

Suppose the following rotating string ansatz:
\begin{eqnarray}
&&t=\kappa\tau\,,\quad\psi_1=\omega\tau\,,\quad\rho=\rho(\sigma)\,,\quad \psi_2=\zeta=0\,,\nonumber\\
&&\phi=\phi_1=\phi_2=\theta=\xi=0\,.
\label{rot}
\end{eqnarray}
Here $\kappa$ and $\omega$ are real constant parameters 
and we assume that $\kappa >0$ and $\omega\geq0$\,. 
The function $\rho(\sigma)$ satisfies the periodic boundary condition: 
$\rho(\sigma) = \rho(\sigma +2\pi)$\,. 
Note that (\ref{rot}) is a consistent ansatz and the WZ term vanishes under (\ref{rot}). 

\medskip 

The conserved charges associated with $t$ and $\psi_1$ are given by, respectively,
\begin{eqnarray}
E&=&\sqrt{\lambda}(1+C^2)^{\frac{1}{2}} \,\kappa \int^{2\pi}_{0}\frac{d\sigma}{2\pi}\,\frac{\cosh^2\rho}{1-C^2\sinh^2\rho}\,,
\label{E}\\
S&=& \sqrt{\lambda}(1+C^2)^{\frac{1}{2}}\,\omega\int^{2\pi}_{0}\frac{d\sigma}{2\pi}\,{\sinh^2\rho}\,.
\label{S}
\end{eqnarray}

\medskip

Under the rotating string ansatz (\ref{rot})\,, the equation of motion for $\rho$ is simplified as 
\begin{eqnarray}
\frac{\rho''}{1-C^2\sinh^2\rho}
=\Bigl[\frac{(1+C^2)\kappa^2-C^2\rho'^{2}}{(1-C^2\sinh^2\rho)^2}\,-\omega^2\Bigr]\sinh\rho\cosh\rho\,.
\label{eom-rho}
\end{eqnarray}
The Virasoro constraints impose the following constraint on $\rho$\,:
\begin{eqnarray}
\frac{\rho'^{2}}{1-C^2\sinh^2\rho}=\kappa^2\frac{\cosh^2\rho}{1-C^2\sinh^2\rho}-\omega^2\sinh^2\rho\,.
\label{v}
\end{eqnarray}
This implies the integral of (\ref{eom-rho})\,.
For latter discussion, it is helpful to rewrite (\ref{v}) as 
\begin{eqnarray}
\frac{\rho'^{2}}{1-C^2\sinh^2\rho}=\frac{C^2\omega^2(\sinh^2\rho_0^{(+)}
-\sinh^2\rho)(\sinh^2\rho_0^{(-)}-\sinh^2\rho)}{1-C^2\sinh^2\rho}\,.
\label{eq-rho}
\end{eqnarray}
Here $\rho_0^{(\pm)}$ are extreme values of $\rho$ given by
\begin{eqnarray}
\sinh^2\rho_0^{(\pm)}=\frac{1}{2C^2\omega^2} 
\Bigl( {\omega^2-\kappa^2\pm\sqrt{ (\omega^2-\kappa^2)^2-4C^2\kappa^2\omega^2}}\Bigr)\,. 
\label{cons}
\end{eqnarray}
The subscript $(\pm)$ corresponds to the signature of the square root. 
For later convenience, it is helpful to see the behavior of $\sinh\rho_0^{(\pm)}$ 
in the $C\rightarrow0$ limit with $\kappa$ and $\omega$ fixed:   
\begin{eqnarray}
\sinh\rho_0^{(+)}\sim\frac{1}{C}\rightarrow\infty\,, \qquad
\sinh\rho_0^{(-)}\rightarrow\frac{\kappa}{\sqrt{\omega^2-\kappa^2}}\,.  
\end{eqnarray}
Note that the r.h.s of (\ref{eq-rho}) remains finite and non-zero. 

\medskip

In the following, we are concerned with the behavior of $\rho$ 
by solving the Virasoro constraint (\ref{eq-rho})\,. 
However, we have to take account of constraint conditions on $\kappa$ and $\omega$\,.  
These come from (\ref{cons}) because $\sinh^2\rho_0^{(\pm)}$ is positive and real. 
Then, we examine the behavior of the solution and compute the conserved charges. 

\subsubsection*{Constraints on $\kappa$ and $\omega$}

Let us consider constraints on $\kappa$ and $\omega$\,, which come from (\ref{cons}) 
and the assumptions:\footnote{Actually, the assumptions for $\rho_0^{(-)}$ are enough. 
One can ensure that $\rho_0^{(+)}$ is real and non-negative from the assumptions for $\rho_0^{(-)}$\,.}  
\begin{eqnarray}
\sinh^2\rho_0^{(\pm)}\in\mathbb{R}\qquad \textrm{and}\qquad \sinh^2\rho_0^{(\pm)}\geq0\,.
\end{eqnarray}
From the reality condition, we obtain an inequality, 
\begin{eqnarray}
(\omega^2-\kappa^2)^2-4C^2\kappa^2\omega^2\geq0\,.
\label{former}
\end{eqnarray}
The positivity leads to other inequalities,  
\begin{eqnarray}
\omega^2-\kappa^2\pm\sqrt{ (\omega^2-\kappa^2)^2-4C^2\kappa^2\omega^2}\geq0\,,
\label{latter}
\end{eqnarray}
where the sign of the square root correspond to $\rho_0^{(\pm)}$\,.
Because $C$ is non-negative, the former condition (\ref{former}) means that 
\begin{eqnarray}
\omega&\geq&(\sqrt{1+C^2}+C)\kappa\qquad\textrm{for}\qquad\omega\geq\kappa\,,
\label{w>k}\\
0\leq\omega&\leq&(\sqrt{1+C^2}-C)\kappa\qquad\textrm{for}\qquad\omega<\kappa\,.
\label{w<k}
\end{eqnarray}
On the other hand, the latter one (\ref{latter}) is satisfied only by the $\omega\geq\kappa$ case. 
For both $\rho^{(\pm)}$\,, the following condition should be satisfied,  
\begin{eqnarray}
\omega&\geq&(\sqrt{1+C^2}+C)\kappa\qquad\textrm{for} \qquad\omega\geq\kappa\,.
\label{latter-w>k}
\end{eqnarray}
As a result, for both $\rho_0^{(\pm)}$\,, just a single inequality is necessary to be satisfied,   
\begin{eqnarray}
\omega\geq(\sqrt{1+C^2}+C)\kappa\,.
\label{w-1}
\end{eqnarray}
This inequality plays a crucial role in studying the behavior of the solution.

\subsubsection*{The range  of $\rho_0^{(\pm)}$}

Let us next examine the range  of $\rho_0^{(\pm)}$ under the condition (\ref{w-1})\,.
Note that the $q$-deformed AdS$_5\times$S$^5$ metric (\ref{ads5}) 
has the singularity at $\rho = \rho_s$\,,  
\begin{eqnarray}
\rho_s=\textrm{arcsinh}\, \frac{1}{C}\,.
\end{eqnarray}
The condition (\ref{w-1}) lead to a sequences of inequalities, 
\begin{eqnarray}
\sinh^2\rho_0^{(-)}  \leq \sinh^2  \rho^*_0 =\sqrt{1+\frac{1}{C^2}}-1
\leq \sinh^2 \rho_0^{(+)} <\sinh^2\rho_s=\frac{1}{C^2}\,.
\label{rho-cond}
\end{eqnarray}
Here $\rho^*_0$ is the value of $\rho$ when $\rho_0^{(-)}=\rho_0^{(+)}$\,, 
which is realized by adjusting $\omega$ and $\kappa$ like 
\begin{eqnarray}
{\omega}=(\sqrt{1+C^2}+C)\kappa\,,
\end{eqnarray}
That is, it is when the inequality (\ref{w-1}) is saturated
and notably the resulting value of  $\rho^*_0$ depends only on $C$\,.
It should be remarked that $\rho^{(\pm)}$ cannot exceed the singularity surface located at $\rho=\rho_s$\,. 
This fact will be significant in later analysis.

\subsubsection*{Allowed regions for the classical solutions}
With the range  of $\rho_0^{(\pm)}$ in (\ref{rho-cond})\,, the positivity of the r.h.s.\ of (\ref{eq-rho}) implies that the solution is allowed to exist only in two regions:
\begin{eqnarray}
&\textrm{I})&\quad0\leq\rho\leq\rho_0^{(-)} \qquad \textrm{with}\qquad \rho_0^{(-)}  \leq \rho^*_0< \rho_s  \,,\\
&\textrm{II})&\quad  \rho_0^{(+)}\leq \rho\leq \infty \qquad \textrm{with}\qquad\rho^*_0\leq \rho_0^{(+)}<\rho_s\,.
\end{eqnarray}
As a result,  a GKP-like string solution can exist only in the region I), because it should contain $\rho=0$\,. 
Thus the solution cannot stretch beyond the singularity surface.

\medskip 

On the other hand, in  the region II)\,, it seems that  spiky string solutions might exist. 
Naively, one might think that the spiky strings stretch from $\rho=\infty$ to $\rho_0^{(+)}$\,, 
passing through the singularity surface, then turn back to $\rho=\infty$\,. 
In this case, however, the integrand of the energy (\ref{E}) 
diverges at the singularity and becomes negative for $\rho>\rho_s$\,.
It may be possible to consider a spiky string solution attached to the singularity surface, 
rather than $\rho=\infty$\,. We will not discuss it here, but it would be quite interesting to 
study it elaborately.

\subsubsection*{The periodicity condition}

It is turn to examine the periodicity of $\rho$ in the region I)\,. 
A folded string satisfies the periodicity by splitting the range $0\leq\sigma<2\pi$  into four segments\,:
for $0 \leq \sigma < \pi/2$ the function $\rho(\sigma)$ increases from zero to its maximal value $\rho_0^{(-)}$\,,
then  $\rho(\sigma)$ decrease to zero for $\pi/2 \leq \sigma < \pi$\,.
The same procedure is followed for $ \pi \leq\sigma<2\pi$ as well.

\medskip

By rewriting $d\sigma$ with (\ref{eq-rho})\,, the periodicity condition can be expressed as 
\begin{eqnarray}
2\pi&=&\int^{2\pi}_{0}\!\!d\sigma=4\int^{\rho_0^{(-)}}_{0}\!\!\frac{d\rho}{\rho'}
=4\int^{\rho_0^{(-)}}_{0}\!\!\!\!\frac{d\rho\,\sinh\rho_0^{(+)}\sinh\rho_0^{(-)}}{\kappa\sqrt{(\sinh^2\rho_0^{(+)}
-\sinh^2\rho)(\sinh^2\rho_0^{(-)}-\sinh^2\rho)}}\nonumber\\
&=&4\int^{\rho_0^{(-)}}_{0}\!\!\!\!\frac{d\rho\,\sinh\rho_0^{(-)}}{\kappa\sqrt{(1-\frac{C^2\omega^2}{\kappa^2}
\sinh^2\rho_0^{(-)}\sinh^2\rho)(\sinh^2\rho_0^{(-)}-\sinh^2\rho)}}\,.
\label{dr}
\end{eqnarray}
Here we have used the relation,
\begin{eqnarray}
\sinh\rho_0^{(+)}\sinh\rho_0^{(-)}=\frac{\kappa}{C\omega}\,.
\end{eqnarray}

\subsubsection*{The conserved charges}

Finally, we evaluate the conserved charges $E$ and $S$\,.  
By using (\ref{dr}), the charges given in (\ref{E}) and (\ref{S}) are rewritten as, respectively,
\begin{eqnarray}
E&=&\sqrt{\lambda}\,(1+C^2)^{\frac{1}{2}}\, \frac{2\kappa}{\pi} \int^{\rho_0^{(-)}}_{0}
\frac{d\rho}{\rho'}\,\frac{\cosh^2\rho}{1-C^2\sinh^2\rho}\,, \nonumber \\
S&=& \sqrt{\lambda}\,(1+C^2)^{\frac{1}{2}}\, \frac{2\omega}{\pi}\int^{\rho_0^{(-)}}_{0}
\frac{d\rho}{\rho'}\,{\sinh^2\rho}\,. \nonumber 
\end{eqnarray}
Note that the charges are not independent of each other 
due to the Virasoro constraints. Therefore the energy $E$ can be expressed as a function of $S$\,. 
However, it is quite difficult to obtain the explicit form of the function in general. 
Hence one needs to take some special limits to study approximate forms. 
In the next subsection, we will consider two kinds of limits and consider the explicit expressions of $E$ 
in terms of $S$\,.

\subsection{Two limits of the solution}

We shall study two kinds of limits of the solution, 1) a short-string limit and 2) a long-string limit.  
In each limit, the classical energy $E$ is expressed as an expansion in terms of the angular momentum $S$\,. 

\subsubsection*{1) a short-string limit}

Let us consider a case that $\rho_0^{(+)}$ is close to the location of the singularity surface $\rho=\rho_s$\,.
For large $\omega$\,, i.e. $\omega \gg \kappa$\,, 
$\sinh\rho_0^{(\pm)}$ behaves like 
\begin{eqnarray}
\sinh^2\rho_0^{(-)}\sim\Bigl(\frac{\kappa}{\omega}\Bigr)^2\,,\qquad \sinh^2\rho_0^{(+)}\sim\frac{1}{C^2}\,.
\end{eqnarray}
This indicates that the solution is confined a narrow region near the origin of the deformed AdS$_5$\,. 
As a result, the length of the folded string is quite short. 

\medskip 

Then, by using approximations $\sinh\rho\approx\rho$ and $\rho_0^{(-)}\approx{\kappa}/{\omega}$\,, 
the conserved charges are evaluated as
\begin{eqnarray}
E&\simeq&\frac{2\sqrt{\lambda}}{\pi}\,\rho_0^{(-)}\!\! \int^{\rho_0^{(-)}}_{0}
\!\!{d\rho}\,\frac{(1+\rho^2)(1+C^2\rho^2)}{\sqrt{(\rho_0^{(-)})^2-\rho^2}}\nonumber\\
&=&\sqrt{\lambda}(1+C^2)^{\frac{1}{2}}\, \frac{\kappa}{\omega}\Bigl[ 1+\frac{1}{2}(1+C^2)
\Bigl(\frac{\kappa}{\omega}\Bigr)^2+...\Bigr]\,,\\
S&\simeq&\frac{2\sqrt{\lambda}}{\pi}\,\frac{\omega}{\kappa}\rho_0^{(-)}\!\! 
\int^{\rho_0^{(-)}}_{0}\!\!{d\rho}\,\frac{\rho^2}{\sqrt{(\rho_0^{(-)})^2-\rho^2}}\nonumber\\
&=& \frac{\sqrt{\lambda}}{2}(1+C^2)^{\frac{1}{2}}\, \Bigl(\frac{\kappa}{\omega}\Bigr)^2+...\,.
\end{eqnarray}
As $\kappa/\omega\ll1$\,, it implies $S/\sqrt{\lambda}\ll1$\,.
As a result, we obtain the following relation
\begin{eqnarray}
E^2=2\sqrt{\lambda}(1+C^2)^{\frac{1}{2}}\,S\Bigl[ 1+(1+C^2)^{\frac{1}{2}}\frac{2S}{\sqrt{\lambda}}+...\Bigr]\,.
\label{short-abf}
\end{eqnarray}
The resulting expression is quite similar to the undeformed case \cite{GKP}, up to the $C$-dependent 
coefficients. When $C=0$\,, the result in \cite{GKP} is reproduced precisely.

\subsubsection*{2) a long-string limit} 

Another interesting case is that $\rho_0^{(-)}\approx\rho_0^{*}$\,, 
where the length of the folded string becomes maximal. 
It is convenient to introduce new variables like
\begin{eqnarray}
\sin u\equiv \frac{\sinh\rho}{~\sinh\rho_0^{(-)}}\,,\qquad q\equiv \sinh^2\rho_0^{(-)}>0\,. \nonumber 
\end{eqnarray} 
Solving $\omega/\kappa$ in terms of $q$\,, we obtain that 
\begin{eqnarray}
\frac{\omega}{\kappa}=\frac{1}{\sqrt{1-C^2q}}\sqrt{\frac{1+q}{q}}\,.
\end{eqnarray}
Then the periodicity condition (\ref{dr}) can be rewritten as
\begin{eqnarray}
2\pi=\frac{4\sqrt{q}}{\kappa}\int^{\frac{\pi}{2}}_{0}\!\!\frac{du}{\sqrt{1+q\sin^2u}}\frac{1}{\sqrt{1-\frac{C^2q(1+q)}{1-C^2q}\sin^2u}}\,.
\label{du}
\end{eqnarray}
In the present case, the following condition is supposed, 
\begin{eqnarray}
\frac{\sinh^2\rho_0^{(-)}}{\sinh^2\rho_0^{(+)}}=\frac{C^2q(1+q)}{1-C^2q}=1-\epsilon\qquad \textrm{with}\qquad \epsilon\ll1\,.
\label{long}
\end{eqnarray}
The condition (\ref{long}) is rewritten as 
\begin{eqnarray}
q\equiv\sqrt{\Bigl(1+\frac{1}{C^2}\Bigr)(1-\epsilon)+\frac{\epsilon^2}{4} }-1+\frac{\epsilon}{2}\qquad \textrm{with}\qquad \epsilon\ll1\,.
\label{eq-k}
\end{eqnarray}
This implies that
\begin{eqnarray}
 \sinh^2\rho_0^{(-)}\equiv q \approx \sqrt{1+\frac{1}{C^2}}-1=\sinh^2  \rho^*_0\,.
\end{eqnarray}
Then the periodicity condition (\ref{du}) leads to the Elliptic integral of the first kind
\begin{eqnarray}
2\pi&=&\frac{4}{\kappa}\sqrt{\frac{q}{1+q}}\textrm{K}\Bigl[1-\frac{\epsilon}{1+q}\Bigr]\nonumber\\
&=&\frac{4}{\kappa}\sqrt{ 1-\frac{C}{\sqrt{1+C^2}} }\Bigl[\frac{1}{2}\log\frac{1}{\epsilon}+\textrm{const.}(C)+\mathcal{O}(\epsilon)\Bigr]\,,
\label{periodic-int}
\end{eqnarray}
where $\textrm{const.}(C)$ is a $C$-dependent constant term.
The finiteness of the integral (\ref{periodic-int}) implies the following limit for consistency,
\begin{eqnarray}
\epsilon\ll1\,,\qquad\kappa\gg1\qquad \textrm{with}\qquad \frac{1}{\kappa}\log\frac{1}{\epsilon}:\textrm{fixed}\,.
\label{long-limit}
\end{eqnarray}
Combining this limit to the condition (\ref{eq-k})\,, 
the consistent  condition for a long-string limit becomes
\begin{eqnarray}
&&q\equiv\sqrt{\Bigl(1+\frac{1}{C^2}\Bigr)(1-\epsilon)+\frac{\epsilon^2}{4} }-1+\frac{\epsilon}{2}\,,\nonumber\\
&&\epsilon\ll1\,,\qquad\kappa\gg1\qquad \textrm{with}\qquad \frac{1}{\kappa}\log\frac{1}{\epsilon}:\textrm{fixed}\,.
\label{long-limit2}
\end{eqnarray}
It implies the following relation
\begin{eqnarray}
q=\sqrt{1+\frac{1}{C^2}}-1+\mathcal{O}(\epsilon)\,,\qquad \frac{\omega}{\kappa}= \sqrt{1+C^2}+C+\mathcal{O}(\epsilon^2)\,.
\label{approx}
\end{eqnarray}
Under the condition (\ref{long-limit2})\,, the periodicity (\ref{periodic-int}) becomes
\begin{eqnarray}
2\pi
=\frac{2}{\kappa}\sqrt{ 1-\frac{C}{\sqrt{1+C^2}} }\,\log\frac{1}{\epsilon}\,. \nonumber 
\end{eqnarray}
Note that even in the $C\rightarrow0$ limit, this solution does not reproduce a GKP string solution.

\medskip

It is turn to examine the conserved charges $E$ and $S$\,. 
Under the condition (\ref{long-limit2})\,, the charges are given by 
\begin{eqnarray}
E
&=&\frac{2\sqrt{\lambda}}{\pi}(1+C^2)^{\frac{1}{2}} \sqrt{q}\int^{\frac{\pi}{2}}_{0}\!\!\,\frac{du\,\sqrt{1+q\sin^2u}}{(1-C^2q\sin^2u)\sqrt{1-(1-\epsilon)\sin^2u}}\,, \nonumber \\
S
&=& \frac{2\sqrt{\lambda}}{\pi}(1+C^2)^{\frac{1}{2}}\frac{\omega}{\kappa}\sqrt{q} \int^{\frac{\pi}{2}}_{0}\!\!\frac{du}
{\sqrt{1+q\sin^2u}}\frac{q\sin^2u}{\sqrt{1-(1-\epsilon)\sin^2u}}\,. \nonumber 
\end{eqnarray}
In order to evaluate the charges, it is helpful to divide the integrals as follows
\begin{eqnarray}
E=\int^{\frac{\pi}{2}-2\epsilon^{1/4}}_{0}\!\!\!\!\!\!\!\!du\,f_E(u)
+\int_{\frac{\pi}{2}-2\epsilon^{1/4}}^{\frac{\pi}{2}}\!\!\!\!\!\!du\,f_E(u)\,,\quad
S
=\int^{\frac{\pi}{2}-2\epsilon^{1/4}}_{0}\!\!\!\!\!\!\!\!du\,f_S(u)
+\int_{\frac{\pi}{2}-2\epsilon^{1/4}}^{\frac{\pi}{2}}\!\!\!\!\!\!du\,f_S(u)\,, \nonumber 
\end{eqnarray}
where $f_E(u)$ and $f_S(u)$ denote the integrands of $E$ and $S$, respectively.

\medskip

We first consider the integrals for $0\leq u\leq\frac{\pi}{2}-2\epsilon^{1/4}$ under the condition (\ref{long-limit2})\,.
In this region, $f_E(u)$ and $f_S(u)$ can be expanded in terms of $\epsilon$ like
\begin{eqnarray}
f_E(u) &=& \frac{2\sqrt{\lambda}}{\pi}(1+C^2)^{\frac{1}{2}} \sqrt{q}\,
\Bigl(\,\frac{\sqrt{1+q\sin^2u}}{\cos u\,(1-C^2q\sin^2u)}+\mathcal{O}(\epsilon)\Bigr)\,, 
\nonumber\\
f_S(u) &=& \frac{2\sqrt{\lambda}}{\pi}(1+C^2)^{\frac{1}{2}}\frac{\omega}{\kappa}\sqrt{q}\,
\Bigl(\,\frac{q\sin^2u}{\cos u\,\sqrt{1+q\sin^2u}}+\mathcal{O}(\epsilon)\Bigr)\,. \nonumber 
\end{eqnarray}
Here $q$ and $\omega/\kappa$ are evaluated as in (\ref{approx})\,.
Then the integrals are also expressed as 
\begin{eqnarray}
&& \int^{\frac{\pi}{2}-2\epsilon^{1/4}}_0\!\!\!\!\!\!\!\!du\,f_E(u) \nonumber \\
&=&\frac{2\sqrt{\lambda}}{\pi}\frac{\sqrt{1+C^2}}{C}\biggl(\frac{1}{4}\sqrt{1+\frac{C}{\sqrt{1+C^2}}}\log\Bigl[\frac{{1+C^2}}{C^2\epsilon}\Bigr]  -\textrm{arctanh}\Bigl(1+\frac{C}{\sqrt{1+C^2}}\Bigr)^{-\frac{1}{2}}
\nonumber\\
&&\hspace{3cm}
+\sqrt{1+\frac{C}{\sqrt{1+C^2}}}\Bigl(\frac{2}{3}+\frac{C}{\sqrt{1+C^2}}\Bigr)\epsilon^\frac{1}{2}
+\mathcal{O}(\epsilon)\biggr)\,,\nonumber
\\
&&\int^{\frac{\pi}{2}-2\epsilon^{1/4}}_0\!\!\!\!\!\!\!\!du\,f_E(u) \nonumber \\
&=&\frac{2\sqrt{\lambda}}{\pi}\frac{\sqrt{1+C^2}}{C}\biggl(\frac{1}{4}\sqrt{1-\frac{C}{\sqrt{1+C^2}}}\log\Bigl[\frac{{1+C^2}}{C^2\epsilon}\Bigr]-\textrm{arcsinh}\Bigl(\sqrt{1+\frac{1}{C^2}}-1\Bigr)^{\!\frac{1}{2}}\nonumber\\
&&\hspace{3cm}+\sqrt{1-\frac{C}{\sqrt{1+C^2}}}\Bigl(\frac{2}{3}+\frac{C}{\sqrt{1+C^2}}\Bigr)\epsilon^\frac{1}{2}+\mathcal{O}(\epsilon)\biggr)\,.\nonumber
\end{eqnarray}

\medskip

Let us next check the integrals for $\frac{\pi}{2}-2\epsilon^{1/4}\leq u\leq\frac{\pi}{2}$\,.
Under the condition (\ref{long-limit2})\,, 
both integrals can be evaluated by the middle-point prescription, respectively, 
\begin{eqnarray}
\int_{\frac{\pi}{2}-2\epsilon^{1/4}}^{\frac{\pi}{2}}\!\!du\,f_E(u)&\simeq &2\epsilon^\frac{1}{4}\,  
f_E\bigl(u=\frac{\pi}{2}-\epsilon^\frac{1}{4}\bigr) \\
& =&  \frac{2\sqrt{\lambda}}{\pi}\frac{\sqrt{1+C^2}}{C}
\sqrt{1+\frac{C}{\sqrt{1+C^2}}}\biggl(2
-\Bigl(\frac{5}{3}+\frac{C}{\sqrt{1+C^2}}\Bigr)\epsilon^\frac{1}{2}+\mathcal{O}(\epsilon)\biggl)\,, \nonumber\\
\int_{\frac{\pi}{2}-2\epsilon^{1/4}}^{\frac{\pi}{2}}\!\!du\,f_S(u)&\simeq &2\epsilon^\frac{1}{4}\,  
f_S\bigl(u=\frac{\pi}{2}-\epsilon^\frac{1}{4}\bigr)  \\
& =& \frac{2\sqrt{\lambda}}{\pi}\frac{\sqrt{1+C^2}}{C}
\sqrt{1-\frac{C}{\sqrt{1+C^2}}}\biggl(2
-\Bigl(\frac{5}{3}+\frac{C}{\sqrt{1+C^2}}\Bigr)\epsilon^\frac{1}{2}+\mathcal{O}(\epsilon)\biggl)\,. \nonumber
\end{eqnarray}

\medskip

As a result, the conserved charges $E$ and $S$ can be evaluated  as, respectively,
\begin{eqnarray}
E
&=&\frac{2\sqrt{\lambda}}{\pi}\frac{\sqrt{1+C^2}}{C}\biggl(\frac{1}{4}\sqrt{1+\frac{C}{\sqrt{1+C^2}}}
\log\Bigl[\frac{{1+C^2}}{C^2\epsilon}\Bigr]-\textrm{arctanh}\Bigl(1+\frac{C}{\sqrt{1+C^2}}\Bigr)^{-\frac{1}{2}}
\nonumber\\
&&\hspace{3cm}+\sqrt{1+\frac{C}{\sqrt{1+C^2}}}(2
-\epsilon^\frac{1}{2})+\mathcal{O}(\epsilon)\biggr)\,, \nonumber 
\\
S
&=&\frac{2\sqrt{\lambda}}{\pi}\frac{\sqrt{1+C^2}}{C}\biggl(\frac{1}{4}\sqrt{1-\frac{C}{\sqrt{1+C^2}}}
\log\Bigl[\frac{{1+C^2}}{C^2\epsilon}\Bigr]-\textrm{arcsinh}
\Bigl(\sqrt{1+\frac{1}{C^2}}-1\Bigr)^{\!\frac{1}{2}}\nonumber\\
&&\hspace{3cm}+\sqrt{1-\frac{C}{\sqrt{1+C^2}}}(2
-\epsilon^\frac{1}{2})+\mathcal{O}(\epsilon)\biggr)\,. \nonumber 
\end{eqnarray}
Thus we have obtained the following relation,  
\begin{eqnarray}
E-(\sqrt{1+C^2}+C)S&=&\frac{2\sqrt{\lambda}}{\pi}\frac{\sqrt{1+C^2}}{C}
\biggl[(\sqrt{1+C^2}+C)\textrm{arcsinh}\Bigl(\sqrt{1+\frac{1}{C^2}}-1\Bigr)^{\!\frac{1}{2}}\nonumber\\
&&\hspace{2cm}-\textrm{arctanh}\Bigl(1+\frac{C}{\sqrt{1+C^2}}\Bigr)^{-\frac{1}{2}}+\mathcal{O}(\epsilon)
\biggr]\,.\label{dis1}
\end{eqnarray}
Note that $\mathcal{O}(\epsilon^{1/2})$ terms have been canceled out in this relation.

\medskip 

By following \cite{talk}, it may be helpful to introduce $w_0$ and $k_0$ defined as 
\begin{eqnarray}
w_0\equiv\sqrt{1+\frac{C}{\sqrt{1+C^2}}}\,,\qquad k_0\equiv\sqrt{1-\frac{C}{\sqrt{1+C^2}}}\,.
\end{eqnarray}
Then it is an easy task to show the relations, 
\begin{eqnarray}
\textrm{arcsinh}\Bigl(\sqrt{1+\frac{1}{C^2}}-1\Bigr)^{\!\frac{1}{2}}
&=&\frac{1}{2}\log\Bigl[ \frac{1+k_0}{1-k_0}\Bigr]\,,\nonumber\\
-\textrm{arctanh}\Bigl(1+\frac{C}{\sqrt{1+C^2}}\Bigr)^{-\frac{1}{2}}
&=&\frac{1}{2}\log\Bigl[ \frac{w_0-1}{w_0+1}\Bigr]\,. \nonumber 
\end{eqnarray}
Thus we have obtained that
\begin{eqnarray}
E-\left(\frac{w_0}{k_0}\right)S=\frac{\sqrt{\lambda}}{\pi}\frac{\sqrt{1+C^2}}{C}
\biggl(\frac{w_0}{k_0}\log\Bigl[ \frac{1+k_0}{1-k_0}\Bigr]+\log\Bigl[ \frac{w_0-1}{w_0+1}\Bigr]
+\mathcal{O}(\epsilon)
\biggr)\,.
\label{dis2}
\end{eqnarray}
Note that the $\mathcal{O}(\epsilon)$ terms are expressed in terms of functions of $S$ through 
\begin{eqnarray}
\epsilon\sim \textrm{exp}\Bigl[ -C\sqrt{1+\frac{C}{\sqrt{1+C^2}}}\frac{2\pi}{\sqrt{\lambda}}\,S\Bigr]\,.
\end{eqnarray}
This result agrees with the one presented in \cite{talk}\,.

\medskip 

Finally, it is worth noting the difference between the dispersion relation (\ref{dis1}) (or equivalently (\ref{dis2})) 
and the GKP dispersion relation in the undeformed case. 
The dispersion relation (\ref{dis1}) is not related to the GKP dispersion relation (log S) through the undeformed limit $C \to 0$. 
In particular, the log factor cannot be reproduced. 
A possible reason is that under the condition (\ref{long-limit2}) one can see that 
$\sinh^2 \rho_0^{(-)} \sim \sinh^2 \rho_0^{(+)} \sim 1/C^2$\,.  
Then the r.h.s.\ of (\ref{eq-rho}) behaves $\sim 1/C^2$ and hence becomes infinite in the $C \to 0$ limit.
It indicates that this solution 
is not reduced to the usual GKP string solution in the undeformed limit.

\section{A new coordinate system for $q$-deformed AdS$_5\times$S$^5$}

A remarkable result in the previous section is that 
the GKP-like string solution cannot stretch beyond the singularity surface 
for an arbitrary value of the angular momentum. This behavior may be regarded as 
a sign that the degrees of freedom are confined into the region inside the singularity 
surface. In order to confirm this statement, we have to do much more efforts 
and get a lot of support. However, if it is true, then it suggests that the holographic 
relation may be discussed only inside the singularity surface. 

\medskip 

Therefore, it would worth trying to look for a good coordinate system 
in which only the inside of the singularity surface is well described. 
This viewpoint is somehow similar to the tortoise coordinates for black holes. 
In the following, we will introduce a new coordinate system in which 
the singularity surface is located at infinity of the radial direction.  
Then we consider minimal surfaces with the new coordinate system.

\subsection{A new coordinate system}

Let us try to introduce a new coordinate system in which only the inside 
of the singularity surface is described. 

\medskip

We first introduce a new coordinate $\chi$\,, instead of $\rho$\,, for the deformed AdS$_5$ part:
\begin{eqnarray}
	\dfrac{\cosh\rho}{\sqrt{1-C^2\sinh^2\rho}}\equiv\cosh\chi\,.
\label{coordtransf}
\end{eqnarray}
Note that $\chi$ coincides with $\rho$ when $C=0$\,.  
This transformation maps 
\begin{eqnarray}
\rho ~\in~ \left[0,\textrm{arcsinh}(1/C)\right) 
\qquad \textrm{to}\qquad
\chi ~\in~[0, \infty)\,.
\end{eqnarray}
The transformation (\ref{coordtransf}) implies that 
\begin{eqnarray}
	\sinh^2\rho=\dfrac{\sinh^2\chi}{1+C^2\cosh^2\chi}\,,\qquad \cosh^2\rho=\dfrac{(1+C^2)\cosh^2\chi}{1+C^2\cosh^2\chi}\,,
\end{eqnarray}
and the following relation is also obtained, 
\begin{eqnarray}
	d\rho=\dfrac{(1+C^2)^{\frac{1}{2}}}{1+C^2\cosh^2\chi}\,d\chi\,.
\end{eqnarray}
Then the metric of the deformed AdS$_5$ (\ref{ads5}) can be rewritten as
\begin{eqnarray}
ds^2_{\textrm{AdS}_5} &=& R^2(1+C^2)^{\frac{1}{2}}
\Bigl[ -\cosh^2\chi\,dt^2
+\dfrac{d\chi^2}{1+C^2\cosh^2\chi}\label{ads5new}\\ 
&&\hspace{-1cm}
+\dfrac{(1+C^2\cosh^2\chi)\sinh^2\chi}{(1+C^2\cosh^2\chi)^2+C^2\sinh^4\chi\sin^2\zeta}\,(d\zeta^2+\cos^2\zeta\,d\psi_1^2)
+\dfrac{\sinh^2\chi\sin^2\zeta\,d\psi_2^2}{1+C^2\cosh^2\chi}\Bigr]\,,\nonumber
\end{eqnarray}
Note that  a curvature singularity exists at $\chi=\infty$ in the new coordinate system as well.
In Appendix A, it is shown that it takes infinite affine time for a massless particle  to reach the singularity. 
Thus the proper distance from any point to the singularity is infinite. 
On the other hand, it does not take infinite time to reach the singularity in the coordinate time. 
This is the same result as in the usual AdS space with the global coordinates. Thus it seems likely to treat the singularity surface as 
the boundary in the holographic setup for the $q$-deformed geometry.

\medskip

Similarly, one may also introduce a new coordinate $\gamma$ for the deformed S$^5$~:
\begin{eqnarray}
	\dfrac{\cos\theta}{\sqrt{1+C^2 \sin^2\theta}}\equiv\cos\gamma\,.
\label{coordtransf2}
\end{eqnarray}
Note that $\gamma$ coincide with $\theta$ when $C=0$\,.  
This transformation maps 
\begin{eqnarray}
\theta ~\in~ [0, \frac{\pi}{2}) \qquad \textrm{to} \qquad
\gamma ~\in~ [0, \frac{\pi}{2})\,.
\end{eqnarray}
The transformation (\ref{coordtransf2}) implies that 
\begin{eqnarray}
	\sin^2\theta=\dfrac{\sin^2\gamma}{1+C^2\cos^2\gamma}\,,
\qquad \cos^2\theta=\dfrac{(1+C^2)\cos^2\gamma}{1+C^2\cos^2\gamma}\,,
\end{eqnarray}
and the following relation is also obtained,  
\begin{eqnarray}
	d\theta=\dfrac{(1+C^2)^{\frac{1}{2}}}{1+C^2\cos^2\gamma}\,d\gamma\,.
\end{eqnarray}
Then the resulting metric of the deformed S$^5$ (\ref{s5}) is given by 
\begin{eqnarray}
ds^2_{\textrm{S}^5} &=& R^2(1+C^2)^{\frac{1}{2}}
\Bigl[ \cos^2\gamma\,d\phi^2
+\dfrac{d\gamma^2}{1+C^2\cos^2\gamma}\label{s5new}\\ 
&&\hspace{.cm}
+\dfrac{(1+C^2\cos^2\gamma)\sin^2\gamma}{(1+C^2\cos^2\gamma)^2
+C^2\sin^4\gamma\sin^2\xi}\,(d\xi^2+\cos^2\xi\,d\phi_1^2)
+\dfrac{\sin^2\gamma\sin^2\xi\,d\phi_2^2}{1+C^2\cos^2\gamma}\Bigr]\,.
\nonumber
\end{eqnarray}
Here an advantage of the new coordinate system is not obvious, 
but it will be appreciated when we study a deformed NR model 
in Sec.\ 4.

\medskip 

Finally, the WZ terms are also rewritten as, respectively,  
\begin{eqnarray}
	\mathcal{L}_{\textrm{AdS}}^{\rm WZ} &=& \frac{T_{(C)}}{2}\,C\,
\epsilon^{\mu\nu}\dfrac{\sinh^4\chi\sin2\zeta}{(1+C^2\cosh^2\chi)^2
+C^2\sinh^4\chi\sin^2\zeta}\,\partial_\mu\psi_1\partial_\nu\zeta\,, 
\label{wz_ads_new} \\  
	\mathcal{L}_{\textrm{S}}^{\rm WZ} &=& -\frac{T_{(C)}}{2}\,C\,
\epsilon^{\mu\nu}\dfrac{\sin^4\gamma\sin2\xi}{(1+C^2\cos^2\gamma)^2
+C^2\sin^4\gamma\sin^2\xi}\,\partial_\mu\phi_1\partial_\nu\xi\,.
\label{wz_s_new}
\end{eqnarray}
Note that the resulting WZ terms are non-singular. 

\medskip 

With the new coordinate system, one may revisit the GKP-like string solution. 
The behavior and the relation of the conserved charges should be identical 
just because the coordinate transformation has been performed.

\subsection{Minimal surfaces}

It would be worth considering minimal surfaces in the new coordinate system. 
The gauge-theory dual has not been unveiled yet, but at least when $C=0$\,, 
these solutions correspond to Wilson loops in the $\mathcal{N}$=4 super Yang-Mills theory. 
The minimal surfaces may be a good clue to seek for the dual gauge theory.   

\subsubsection*{1) a static string solution}

Let us first study a static configuration of the string world-sheet 
with the ansatz\,:
\begin{eqnarray}
&&t=\kappa\tau\,,\qquad 
\chi=\chi(\sigma)\,,
\qquad\zeta=\psi_1=\psi_2=0\,,\nonumber\\ &&
\phi=\phi_1=\phi_2=\gamma=\xi=0\,.
\label{ansatz-wilson2}
\end{eqnarray}
Then the full metric is reduced to that of a deformed AdS$_2$ subspace,
\begin{eqnarray}
ds^2_{\textrm{AdS}_2} &=& R^2(1+C^2)^{\frac{1}{2}}
\Bigl[- \cosh^2\chi dt^2+\dfrac{d\chi^2}{1+C^2\cosh^2\chi}
\Bigr]\,.\label{ds-wilson2}
\end{eqnarray}
The WZ terms vanish under this ansatz (\ref{ansatz-wilson2}) and 
the string Lagrangian is given by 
\begin{eqnarray}
\mathcal{L}_{\textrm{AdS}_2} &=& -\frac{\sqrt{\lambda}}{4\pi}\,(1+C^2)^{\frac{1}{2}}
\Bigl[\kappa^2\cosh^2\chi +\dfrac{\chi'^2}{1+C^2\cosh^2\chi}
\Bigr]\,.\label{L-wilson2}
\end{eqnarray}
Then the equation of motion for $\chi$\,, 
\begin{eqnarray}
\frac{\chi''}{1+C^2\cosh^2\chi}&=&\Bigl[\kappa^2
+\frac{C^2\chi'^2}{(1+C^2\cosh^2\chi)^2}\Bigr]\cosh\chi\sinh\chi \,,
\end{eqnarray}
can be integrated as follows: 
\begin{eqnarray}
\kappa^2\cosh^2\chi -\dfrac{\chi'^2}{1+C^2\cosh^2\chi} =A \quad \mbox{(const.)}\,.
\end{eqnarray}
Note that the Virasoro constraint 
\begin{eqnarray}
0=-\kappa^2\cosh^2\chi +\dfrac{\chi'^2}{1+C^2\cosh^2\chi}\,,
\label{vir-wilson2}
\end{eqnarray}
leads to the condition $A=0$\,.

\medskip

Let us solve the first-order differential equation\footnote{Here the $(-)$-signature is picked up, 
but the result is similar for the $(+)$-signature. 
The difference is that the classical solution is stretched from the boundary for $(-)$ or the origin for $(+)$. 
Here we are interested in the solution ending on the boundary, hence we have taken the $(-)$-signature. }, 
\begin{eqnarray}
\chi'&=& -\kappa\cosh\chi\sqrt{1+C^2\cosh^2\chi}\,,
\label{chi-wilson}
\end{eqnarray}
The other direction of the AdS part can be described by the negative $\sigma$ region. 
In total, the solution stretches from one boundary to the other boundary through the center. 

\medskip 

By integrating (\ref{chi-wilson}) with the boundary condition $\chi(\sigma=0)=\infty$\,, 
 \begin{eqnarray}
- \int_\infty^\chi  \frac{d\tilde{\chi}}{\cosh\tilde{\chi}\sqrt{1+C^2\cosh^2\tilde{\chi}}}= \kappa \int_0^\sigma\!\!d\tilde{\sigma}\,,
 \end{eqnarray}
the following expression is obtained,  
\begin{eqnarray}
\kappa\sigma=\textrm{arctan}\Bigl[\frac{1}{C}\Bigr]-\textrm{arctan}\Bigl[\frac{\sinh\chi}{\sqrt{1+C^2\cosh^2\chi}}\Bigr]\,. 
\label{sol21}
\end{eqnarray}
By introducing $\sigma_0$ as $\sinh\chi(\sigma_0) =0$\,,
$\kappa\sigma_0$ is related to $C$ through 
\begin{eqnarray}
\kappa\sigma_0=\textrm{arctan}\Bigl[\frac{1}{C}\Bigr]\,. \label{tan}
\end{eqnarray}
The solution (\ref{sol21}) can be rewritten as
\begin{eqnarray}
\cosh^2\chi=\frac{1+\tan^2\bigl[\textrm{arctan}
[\frac{1}{C}]-\kappa\sigma\bigr]}{1-C^2\tan^2\bigl[\textrm{arctan}[\frac{1}{C}]-\kappa\sigma\bigr]}\,.
\label{straight-sol}
\end{eqnarray}
Then, after taking acount of the negative $\sigma$ region, the total energy is given by
\begin{eqnarray}
E&=& 2 \times \frac{\sqrt{\lambda}}{2\pi}\,(1+C^2)^\frac{1}{2}\,\kappa\int_\epsilon^{\sigma_0}
\!\!d\sigma\cosh^2\chi\nonumber\\
&=&\frac{\sqrt{\lambda}}{\pi}  \frac{\sqrt{1+C^2}}{C}\,\textrm{arctanh}\Bigl[
C\cot\bigl(\kappa\epsilon+\textrm{arctan}[C]\bigr)
\Bigr]\,, 
\label{24}
\end{eqnarray}
where $\epsilon$ is a UV cut-off. 
When $C \neq 0$\,,
one can expand (\ref{24}) in terms of $\epsilon$\,,
\begin{eqnarray}
E=\frac{\sqrt{\lambda}}{\pi}\, \frac{\sqrt{1+C^2}}{C}\,\log\left[\frac{2C}{(1+C^2)\,\kappa\epsilon} \right]+\mathcal{O}(\epsilon)\,.
\end{eqnarray}
Note that the energy $E$ diverges logarithmically unlike the usual AdS/CFT case. 
It does not seem possible to remove this divergence by following the standard procedure \cite{DGO} 
with the Poincare coordinates, as we will show later. 
There might be a proper method to regularize (\ref{24}), or it may be divergent essentially. 






\subsubsection*{A boundary term in Poincare coordinates}

It would be worth mentioning about an additional contribution which comes from the boundary \cite{DGO}.
For that purpose, it is helpful to convert 
the solution (\ref{straight-sol}) in the global coordinates 
into the one in the associated Poinacre coordinates in \cite{future}. 
The metric of the $q$-deformed AdS$_5$ (\ref{ads5new}) can be rewritten with the Poincare coordinates like 
\cite{future}
\begin{eqnarray}
ds^{2}_{\textrm{AdS}_5}&=& R^2\sqrt{1+C^2}
\biggl[ \frac{dz^2+dr^2}{z^2+C^2(z^2+r^2)}
+ \frac{C^2(z\,dz+r\,dr)^2}{z^2\bigl(z^2+C^2(z^2+r^2)\bigr)}
\label{poincare}
\\ 
&&\hspace{-1cm}
+ \frac{\bigl(z^2+C^2(z^2+r^2)\bigr)r^2}{
\bigl(z^2+C^2(z^2+r^2)\bigr)^2+C^2r^4\sin^2\zeta}\,(d\zeta^2+\cos^2\zeta\,d\psi_1^2 ) 
+  \frac{r^2\sin^2\zeta\,d\psi_2^2}{z^2+C^2(z^2+r^2)}
\biggr]
\,.\nonumber
\end{eqnarray}
The solution (\ref{straight-sol}) in the global coordinates can be rewritten in terms of the Poincare coordinates 
and it is divided into two sections according to the positive (or negative) $\sigma$ region. 
For the positive $\sigma$ region, the solution is given by 
\begin{eqnarray}
z^{(+)}(\tau,\sigma)&=&\textrm{e}^{\kappa\tau}\sqrt{\cos^2(\textrm{arctan}[1/C]-\kappa\sigma)-C^2\sin^2(\textrm{arctan}[1/C]-\kappa\sigma)}\,,\nonumber\\
r^{(+)}(\tau,\sigma)&=&\sqrt{1+C^2}\,\textrm{e}^{\kappa\tau}\sin(\textrm{arctan}[1/C]-\kappa\sigma)\,,
\label{zr+}
\end{eqnarray}
and, for the negative $\sigma$ region, the one is  
\begin{eqnarray}
z^{(-)}(\tau,\sigma)&=&\textrm{e}^{\kappa\tau}\sqrt{\cos^2(\textrm{arctan}[1/C]+\kappa\sigma)-C^2\sin^2(\textrm{arctan}[1/C]+\kappa\sigma)}\,,\nonumber\\
r^{(-)}(\tau,\sigma)&=&-\sqrt{1+C^2}\,\textrm{e}^{\kappa\tau}\sin(\textrm{arctan}[1/C]+\kappa\sigma)\,.
\label{zr-}
\end{eqnarray}
Here  $\tau$ is taken as $-\infty<\tau<\infty$ and the $(\pm)$-signatures correspond to the solution in the positive $\sigma$ region $\bigl(0\leq\sigma\leq\sigma_0\bigr)$ and that of the negative $\sigma$ region $\bigl(-\sigma_0\leq\sigma<0)$, respectively.
The positive $\sigma$ region is mapped to $\theta\equiv \textrm{arctan}(r/z)$\,,
\begin{eqnarray}
\theta(\sigma=0)=\frac{\pi}{2}\,~~(\textrm{boundary})\,~~\longrightarrow\,~~\theta(\sigma=\sigma_0)=0 \,~~(\textrm{origin})\,,
\end{eqnarray}
while the negative $\sigma$ region covers
\begin{eqnarray}
\theta(\sigma=-\sigma_0)=0\,~~(\textrm{origin})\,~~\longrightarrow\,~~\theta(\sigma=0)=-\frac{\pi}{2}\,~~(\textrm{boundary})\,.
\end{eqnarray}
Note that the contribution coming from
the negative $\sigma$ region is the same value as that from the positive $\sigma$ region.

\medskip 

Then the classical Euclidean action of this solution is evaluated as
\begin{eqnarray}
S&=&\frac{\sqrt{\lambda}}{2\pi}\sqrt{1+C^2}\int_{-\frac{T}{2}}^{\frac{T}{2}}\!\!d\tau
\int^{\sigma_0}_\epsilon\!\!d\sigma
\,2\kappa^2\Bigl(\frac{1+\tan^2\bigl[\textrm{arctan}
[\frac{1}{C}]-\kappa\sigma\bigr]}{1-C^2\tan^2\bigl[\textrm{arctan}[\frac{1}{C}]-\kappa\sigma\bigr]}\Bigr)
\nonumber\\
&=&\frac{\sqrt{\lambda}}{\pi}\, \frac{\sqrt{1+C^2}}{C}\,T\,\kappa\,\textrm{arctanh}\Bigl[
C\cot\bigl(\kappa\epsilon+\textrm{arctan}[C]\bigr)
\Bigr]\,,
\end{eqnarray}
where $T$ is the interval of the world-sheet $\tau$ and taken to be large. 
When $C \neq 0$\,, the above expression can be expanded in terms of $\epsilon$ like 
\begin{eqnarray}
S=\frac{\sqrt{\lambda}}{\pi}\, \frac{\sqrt{1+C^2}}{C}\,T\,\kappa\,\log\left[\frac{2C}{(1+C^2)\,\kappa\epsilon} \right]+\mathcal{O}(\epsilon)\,.
\label{SE}
\end{eqnarray}
Note that the divergence is logarithmic again, unlike the usual AdS/CFT case.


\medskip

It is turn to consider an additional contribution which comes from the boundary \cite{DGO}.
By taking account of a Legendre transformation, the total action is written into the following form, 
\begin{eqnarray}
\tilde{S}&=&S+S_L\,,\qquad
S_L=2
\int_{-\frac{T}{2}}^{\frac{T}{2}}\!\!d\tau\int^{\sigma_0}_\epsilon\!\!d\sigma\,\partial_\sigma\Bigl(z\,\frac{\partial L}{\partial(\partial_\sigma z)}\Bigr)\,.
\label{S-straightL}
\end{eqnarray}
The term $S_L$ is evaluated as
\begin{eqnarray}
S_L &=& 2\int_{-\frac{T}{2}}^{\frac{T}{2}}\!\!d\tau\Bigl(z\,\frac{\partial L}{\partial(\partial_\sigma z)}\Bigl|_{\sigma_0}-z\,\frac{\partial L}{\partial(\partial_\sigma z)}\Bigl|_{\epsilon}\,\Bigr) \nonumber \\ 
&=&-\frac{\sqrt{\lambda}}{\pi}\,\sqrt{1+C^2}\,T\,\kappa\,\cot\bigl(\kappa\epsilon+\textrm{arctan}[C]\bigr)\,.\nonumber
\end{eqnarray}
When $C \neq 0$\,, $S_L$ is expanded in terms of $\epsilon$ with non-vanishing $C$ like 
\begin{eqnarray}
S_L &=& -\frac{\sqrt{\lambda}}{\pi}\,\frac{\sqrt{1+C^2}}{C}\,\kappa+\mathcal{O}(\epsilon)\,.
\label{SL}
\end{eqnarray}
Thus, even though the boundary contribution (\ref{SL}) has been taken into account, 
the divergent term in (\ref{SE}) cannot be canceled out.

\medskip

It is worth noting the undeformed limit of the classical action (\ref{SE}). 
By taking the $C\rightarrow 0$ limit ($\epsilon$\,:\,fixed), 
the well-known result in the undeformed case is reproduced, 
\begin{eqnarray}
S=\frac{\sqrt{\lambda}}{\pi}\,T\,\kappa\,\textrm{cot}[\kappa\epsilon]=\frac{\sqrt{\lambda}\,T}{\pi\,\epsilon}+\mathcal{O}(\epsilon)\,.
\label{c0}
\end{eqnarray}
In the undeformed limit $C\to 0$ with $\epsilon$ fixed, the expression (\ref{SL}) results in 
\begin{eqnarray}
S_L=-\frac{\sqrt{\lambda}}{\pi}\,T\,\kappa\,\textrm{cot}[\kappa\epsilon]\,,
\end{eqnarray}
and it cancels out the divergent term in (\ref{c0}) as usual.


\subsubsection*{2)  rotating minimal surface}

We next generalize the minimal surface solution 
by adding an angular momentum in S$^5$\,. 
This is also a generalization of the solution in \cite{drukker} 
to the deformed case. 

\medskip 

Let us start from the following ansatz\,:
\begin{eqnarray}
&&t=\kappa\tau\,,\qquad 
\chi=\chi(\sigma)\,,
\qquad\zeta=\psi_1=\psi_2=0\,,\nonumber\\ &&
\phi=\omega\tau\,,\qquad 
\gamma=\gamma(\sigma)\,,\qquad\xi=\phi_1=\phi_2=0\,.
\label{ansatz-wilson}
\end{eqnarray}
Then the WZ terms vanish under this ansatz and the Lagrangian is expressed as 
\begin{eqnarray}
\mathcal{L}_{\textrm{AdS}_2\times\textrm{S}^2} &=& -\frac{\sqrt{\lambda}}{4\pi}\,(1+C^2)^{\frac{1}{2}}
\Bigl[\kappa^2\cosh^2\chi +\dfrac{\chi'^2}{1+C^2\cosh^2\chi} -\omega^2\cos^2\gamma
+\dfrac{\gamma'^2}{1+C^2\cos^2\gamma}
\Bigr]\,. \nonumber 
\end{eqnarray}
The equations of motion are given by 
\begin{eqnarray}
\frac{\chi''}{1+C^2\cosh^2\chi}&=&\Bigl[\kappa^2+\frac{C^2\chi'^2}{(1+C^2\cosh^2\chi)^2}
\Bigr]\cosh\chi\sinh\chi \,, 
\nonumber \\
\frac{\gamma''}{1+C^2\cos^2\gamma}&=&\Bigl[\omega^2
-\frac{C^2\gamma'^2}{(1+C^2\cos^2\gamma)^2}\Bigr]\cos\gamma\sin\gamma\,. 
\nonumber 
\end{eqnarray}
The Virasoro constraint is 
\begin{eqnarray}
0=-\kappa^2\cosh^2\chi +\dfrac{\chi'^2}{1+C^2\cosh^2\chi}+\omega^2\cos^2\gamma
+\dfrac{\gamma'^2}{1+C^2\cos^2\gamma}\,.
\label{vir-wilson}
\end{eqnarray}
The two equations of motion can be integrated, respectively,
then the Virasoro constraint (\ref{vir-wilson}) 
imposes that the two integrals of motion should be identical, 
\begin{eqnarray}
\kappa^2\cosh^2\chi -\dfrac{\chi'^2}{1+C^2\cosh^2\chi}=\omega^2\cos^2\gamma
+\dfrac{\gamma'^2}{1+C^2\cos^2\gamma}=A\,,
\end{eqnarray}
where $A$ is a common integral of motion.

\medskip

Imposing $\omega=\kappa$ and $A=\kappa^2$\,, 
the classical solution of $\chi$ becomes simple. 
The motion of $\chi$ is fixed by the first-order differential equation\footnote{Here we have chosen the $(-)$-signature due to 
the same reason 
as we have mentioned for the static case. }, 
\begin{eqnarray}
\chi'&=& -\kappa\sinh\chi\sqrt{1+C^2\cosh^2\chi}\,.
\label{chi-wilson2}
\end{eqnarray}
This equation implies that the minimum of $\chi$ is the center of the deformed AdS$_5$ part. 
The other direction of the AdS part is described by the negative $\sigma$ region. 
In total, the solution is stretched from one boundary to the other boundary through the center.

\medskip 

By integrating (\ref{chi-wilson2}) with the boundary condition $\chi(\sigma=0)=\infty$\,,
\begin{eqnarray}
- \int_\infty^\chi  \frac{d\tilde{\chi}}{\sinh\tilde{\chi}\sqrt{1+C^2\cosh^2\tilde{\chi}}}= \kappa \int_0^\sigma\!\!d\tilde{\sigma}\,,
\end{eqnarray}
the following expression is obtained, 
\begin{eqnarray}
\kappa\sigma=\frac{1}{\sqrt{1+C^2}}\,\textrm{arctanh}\Bigl[\frac{\sqrt{1+C^2}\cosh\chi}{\sqrt{1+C^2\cosh^2\chi}}\Bigr]
- \frac{1}{\sqrt{1+C^2}}\,\textrm{arctanh}\Bigl[\frac{\sqrt{1+C^2}}{C}\Bigr]\,. \nonumber 
\end{eqnarray}
This solution can be rewritten as
\begin{eqnarray}
\cosh\chi=\frac{\sinh\bigl[\sqrt{1+C^2}\kappa\sigma+\textrm{arccoth}\,[\frac{C}{\sqrt{1+C^2}}]\bigr]}{
\sqrt{\cosh^2\bigl[\sqrt{1+C^2}\kappa\sigma+\textrm{arccoth}\,[\frac{C}{\sqrt{1+C^2}}]\bigr]+C^2}}\,.
\label{chi-sol-wilson2}
\end{eqnarray}
Taking the $C\rightarrow 0$ limit, the undeformed solution is reproduced, 
\begin{eqnarray}
\cosh\chi\rightarrow\frac{1}{\tanh{\kappa\sigma}}\,. \nonumber 
\end{eqnarray}

\medskip

Next, let us focus upon the sphere part.
The equation for $\gamma$ becomes\footnote{Here we have taken the $(-)$-signature, but the result is similar for the $(+)$-signature. 
The only difference is that the classical solution extends over the northern hemisphere for $(-)$ or southern hemisphere for $(+)$. }
\begin{eqnarray}
\gamma'&=& -\kappa\sin\gamma\,\sqrt{1+C^2\cos^2\gamma}\,.
\label{gamma-wilson}
\end{eqnarray}
By integrating (\ref{gamma-wilson}) with the boundary condition $\gamma(\sigma=0)=\pi/2$\,, the following relation is obtained, 
\begin{eqnarray}
-\int_\frac{\pi}{2}^\gamma \frac{d\tilde{\gamma}}{\sin\tilde{\gamma}\sqrt{1+C^2\cos^2\tilde{\gamma}}}
= \kappa \int_0^\sigma\!\!d\tilde{\sigma}\,.
\label{int-gamma}
\end{eqnarray}
Then the left-hand side of (\ref{int-gamma}) can be rewritten as 
\begin{eqnarray}
\int^\frac{\pi}{2}_\gamma \frac{d\tilde{\gamma}}{\sin\tilde{\gamma}\sqrt{1+C^2\cos^2\tilde{\gamma}}}
=  \frac{1}{\sqrt{1+C^2}}\,\textrm{arctanh}\Bigl[\frac{\sqrt{1+C^2}\cos\gamma}{\sqrt{1+C^2\cos^2\gamma}}\Bigr]\,. 
\nonumber 
\end{eqnarray}
Thus we obtain the solution, 
\begin{eqnarray}
\cos\gamma=\frac{\sinh[\sqrt{1 + C^2} \kappa\sigma] }{ \sqrt{ \cosh^2[\sqrt{1 + C^2} \kappa\sigma] +  C^2 }}\,.
\label{gamma-sol-wilson}
\end{eqnarray}
By taking the $C\rightarrow 0$ limit, the solution (\ref{gamma-sol-wilson}) is reduced to
\begin{eqnarray}
\cos\gamma\rightarrow{\tanh{\kappa\sigma}}\,. \nonumber 
\end{eqnarray}

\medskip

Finally, we examine the conserved charges.
By using the solutions (\ref{chi-sol-wilson2}) and (\ref{gamma-sol-wilson})\,, 
the total angular momentum $J$ and the energy $E$ are given by, respectively,
\begin{eqnarray}
J&=&\frac{\sqrt{\lambda}}{\pi}\, (1+C^2)^\frac{1}{2}\,\kappa\int_0^{\sigma_0}\!\!d\sigma\cos^2\gamma
\nonumber\\
&=&\frac{\sqrt{\lambda}}{\pi}\, (1+C^2)^\frac{1}{2}\,\Bigl(\kappa\sigma_0 - \frac{1}{C}\textrm{arctanh}
\Bigl[\frac{C}{\sqrt{1+C^2}} \tanh[\sqrt{1 + C^2}\kappa\sigma_0]\Bigr]\Bigr)\,, \nonumber \\
E&=&\frac{\sqrt{\lambda}}{\pi}\,  (1+C^2)^\frac{1}{2}\,\kappa\int_\epsilon^{\sigma_0}\!\!d\sigma\cosh^2\chi
\nonumber\\
&=&\frac{\sqrt{\lambda}}{\pi}\,  (1+C^2)^\frac{1}{2}\,\Bigl(\kappa\sigma_0 - \frac{1}{C}\textrm{arctanh}
\Bigl[\frac{C}{\sqrt{1+C^2}}\tanh\bigl[\sqrt{1+C^2}\kappa\sigma_0
+\textrm{arccoth}[\frac{C}{\sqrt{1+C^2}}]\bigr]\Bigr)\nonumber\\
&&\hspace{3.cm}+\frac{1}{C}\textrm{arctanh}\bigl[1-\frac{\kappa}{C}\epsilon\bigr]\Bigr) \,, \nonumber 
\end{eqnarray}
where $\epsilon$ is a cut-off, while $\sigma_0$ is a cutoff for the length of the world-sheet.
In the large $\kappa\sigma_0$ limit, it implies that
\begin{eqnarray}
E=J\,, \nonumber 
\end{eqnarray}
after subtracting the UV divergence. 

\medskip 

By taking the $C\rightarrow 0$ limit, the charges are reduced to 
the undeformed ones \cite{drukker}\,,
\begin{eqnarray}
J\rightarrow\frac{\sqrt{\lambda}}{\pi} \,\Bigl(\kappa\sigma_0 - \tanh\kappa\sigma_0\Bigr)\,,\quad
E\rightarrow\frac{\sqrt{\lambda}}{\pi}  \,\Bigl(\kappa\sigma_0 - \coth\kappa\sigma_0+\coth\kappa\epsilon\Bigr)\,. 
\nonumber 
\end{eqnarray}

\section{NR system from $q$-deformed S$^5$ }

In this section, we derive a deformed Neumann-Rosochatius (NR) system 
by introducing a rigid ansatz. 
As a remarkable feature, the resulting system has modified potential terms expressed 
by trigonometric and hyperbolic functions. 
Here we will concentrate on the Lagrangian for the deformed S$^5$ part (\ref{s5new}). 
For the analysis for the deformed AdS$_5$ part, see Appendix B. 

\medskip 

A deformed Neumann model has been derived in the original 
coordinate system \cite{AM-NR}. Probably, there would be an intimate relation 
between the result in \cite{AM-NR} and our result.

\subsection{Rigid string ansatz}

First of all, let us rewrite the metric (\ref{s5new}) with the radial coordinates $r_i$\,, 
\begin{eqnarray}
 r_1=\sin\gamma\cos\xi\,,\qquad r_2=\sin\gamma\sin\xi\,,\qquad r_3=\cos\gamma\,, \qquad \sum_{1=1}^3r_i^2=1
\,.
\end{eqnarray} 
The resulting metric (\ref{s5new}) is given by  
\begin{eqnarray}
ds^2_{\textrm{S}^5} &=& R^2(1+C^2)^{\frac{1}{2}}
\Bigl[ \dfrac{dr_3^2}{1+C^2r_3^2}+r_3^2\,d\phi^2
+\dfrac{dr_1^2}{1+C^2r_3^2}+\dfrac{(1+C^2r_3^2)r_1^2\,d\phi_1^2}{(1+C^2r_3^2)^2+C^2(r_1^2+r_2^2)r_2^2}\nonumber\\ 
&&\hspace{.cm}+\dfrac{dr_2^2+r_2^2\,d\phi_2^2}{1+C^2r_3^2}-\dfrac{C^2r_2^2(r_1dr_2-r_2dr_1)^2}{(1+C^2r_3^2)[(1+C^2r_3^2)^2+C^2(r_1^2+r_2^2)r_2^2]}
\Bigr]\,.\label{S5}
\end{eqnarray}
The WZ part can also be rewritten as
\begin{eqnarray}
	\mathcal{L}_{\textrm{S}}^{WZ} &=& -\frac{T_{(C)}}{2}\,
\epsilon^{\mu\nu}\dfrac{2C r_1r_2\,\partial_\mu\psi_1(r_1\partial_\nu r_2-r_2\partial_\nu r_1)}{(1+C^2r_3^2)^2+C^2(r_1^2+r_2^2)r_2^2}\,.
\label{WZ_s}
\end{eqnarray}

\medskip 

Let us suppose the following rigid string ansatz : 
\begin{eqnarray}
&&\phi_1=\omega_1\tau+\alpha_1(\sigma)\,,\qquad 
\phi_2=\omega_2\tau+\alpha_2(\sigma)\,,\qquad 
\phi=\omega_3\tau+\alpha_3(\sigma)\,,\nonumber\\ 
&&\gamma=\gamma(\sigma)\,,\qquad\xi=\xi(\sigma)\,.
\label{rigid_s}
\end{eqnarray}
According to this ansatz, the periodicity conditions are translated as follows: 
\begin{eqnarray}
&&r_i(\sigma+2\pi)=r_i(\sigma)\qquad(i=1,2,3)\,, \nonumber \\
&&\alpha_i(\sigma+2\pi)=\alpha_i(\sigma)+2\pi m_i\qquad(m_i\in\mathbb{Z})\,. 
\nonumber 
\end{eqnarray}

\subsection{Reduction of deformed S$^5$ to NR system} 

Under the ansatz (\ref{rigid_s})\,, the Lagrangian results in a one-dimensional dynamical system,
\begin{eqnarray}
\mathcal{L}_{\textrm{S}^5}&=&-\frac{T_{(C)}}{2}
\Bigl[ 
\dfrac{r_3'^{2}}{1+C^2r_3^2}+r_3^2(-\omega_3^2+\alpha_3'^{2})
+\dfrac{r_1'^{2}}{1+C^2r_3^2} 
+\dfrac{(1+C^2r_3^2)r_1^2(-\omega_1^2+\alpha_1^{'2})}{(1+C^2r_3^2)^2+C^2(r_1^2+r_2^2)r_2^2}
\nonumber\\ &&\hspace{1cm}+\dfrac{r_2'^{2}+r_2^2(-\omega_2^2+\alpha_2'^{2})}{1+C^2r_3^2}
-\dfrac{C^2r_2^2(r_1r_2'-r_2r'_1)^2}{(1+C^2r_3^2)[(1+C^2r_3^2)^2+C^2(r_1^2+r_2^2)r_2^2]}
\nonumber\\ &&\hspace{1cm}
+\dfrac{2C\omega_1 r_1r_2(r_1r'_2-r_2r'_1)}{(1+C^2r_3^2)^2+C^2(r_1^2+r_2^2)r_2^2}
-{\Lambda}(\sum_{i=1}^3r_i^2-1)\Bigr]\,,
\label{s5_rigid}
\end{eqnarray} 
where ${\Lambda}$ is  a Lagrange multiplier.
From the equations of motion for $\alpha_i(\sigma)$\,, three constants of motion are obtained,  
\begin{eqnarray}
v_1\equiv\dfrac{(1+C^2r_3^2)r_1^2\alpha_1'}{(1+C^2r_3^2)^2+C^2(r_1^2+r_2^2)r_2^2}\,,
\quad v_2\equiv\dfrac{r_2^2\alpha'_2}{1+C^2r_3^2}\,,\quad 
v_3\equiv r_3^2\alpha'_3\,,\quad
\end{eqnarray} 
where $v_i$ are three integration constants. 
Note that these constants are modified due to the deformation.
When $C=0$\,, $v_i$ are reduced to the undeformed ones, i.e. $r_i^2\alpha^{'}_i$ \cite{AFT-N}\,.

\medskip

By removing $\alpha_i$ from (\ref{s5_rigid})\,, one can derive the Lagrangian for $r_i$\,, 
\begin{eqnarray}
{L}_{\textrm{S}^5}&=&\frac{T_{(C)}}{2}
\Bigl[\dfrac{r_3'^{2}}{1+C^2r_3^2}-r_3^2\omega_3^2- \frac{v_3^2}{r_3^2}
+\dfrac{r_1'^{2}}{1+C^2r_3^2}-\dfrac{(1+C^2r_3^2)r_1^2\omega_1^2}{(1+C^2r_3^2)^2+C^2(r_1^2+r_2^2)r_2^2}
\nonumber\\
&&\hspace{.cm}-\Bigl(1+C^2r_3^2+\frac{C^2(r_1^2+r_2^2)r_2^2}{1+C^2r_3^2}\Bigr)\,\frac{v_1^2}{r_1^2}
+\dfrac{r_2'^{2}-r_2^2\omega_2^2}{1+C^2r_3^2}-(1+C^2r_3^2)\frac{v_2^2}{r_2^2}\nonumber\\ 
&&\hspace{.cm}-\dfrac{C^2r_2^2(r_1r_2'-r_2r'_1)^2}{(1+C^2r_3^2)[(1+C^2r_3^2)^2+C^2(r_1^2+r_2^2)r_2^2]}
+\dfrac{2C\omega_1 r_1r_2(r_1r'_2-r_2r'_1)}{(1+C^2r_3^2)^2+C^2(r_1^2+r_2^2)r_2^2}\nonumber\\
&&\hspace{.cm}-{\Lambda}(\sum_{i=1}^3r_i^2-1)\Bigr]\,.
\label{s5NR}
\end{eqnarray} 
Here we have changed the sign of (\ref{s5_rigid}), except for the terms which include 
$ v_i^2/ r_i^2$ so as to reproduce the same equations of motion for $r_i$\,. 
As a result, the Lagrangian (\ref{s5NR}) can be regarded as a deformed 
NR system derived from the deformed S$^{5}$\,.
When $C=0$\,, this system reduces to the usual NR system \cite{AFT-N}\,.

\subsubsection*{Virasoro constraints}

Let us comment on the Virasoro constraints. By including the time coordinate in the AdS part 
with $t=\kappa\tau$\,, the Virasoro constraints become
\begin{eqnarray}
\kappa^2&=& \dfrac{r_3'^{2}}{1+C^2r_3^2}+r_3^2\omega_3^2+ \frac{v_3^2}{r_3^2}
+\dfrac{r_1'^{2}}{1+C^2r_3^2}+\dfrac{(1+C^2r_3^2)r_1^2\omega_1^2}{(1+C^2r_3^2)^2+C^2(r_1^2+r_2^2)r_2^2}\nonumber\\
&&\hspace{.cm}+\Bigl(1+C^2r_3^2+\frac{C^2(r_1^2+r_2^2)r_2^2}{1+C^2r_3^2}\Bigr)\,\frac{v_1^2}{r_1^2}
+\dfrac{r_2'^{2}+r_2^2\omega_2^2}{1+C^2r_3^2}+(1+C^2r_3^2)\frac{v_2^2}{r_2^2}\nonumber\\ 
&&\hspace{.cm}-\dfrac{C^2r_2^2(r_1r_2'-r_2r'_1)^2}{(1+C^2r_3^2)[(1+C^2r_3^2)^2+C^2(r_1^2+r_2^2)r_2^2]}\,,\label{v1}\\  
0&=&\sum_{i=1}^3\omega_iv_i\,.\label{v2}
\end{eqnarray}
The second condition (\ref{v2}) implies that only two of  $v_i$ are independent of $\omega_i$\,.
It is also necessary to impose the periodicity conditions for $\alpha_i(\sigma)$:
\begin{eqnarray}
2\pi m_1&=&v_1\int_0^{2\pi}\frac{d\sigma}{r_1^2(\sigma)}\Bigl(1+C^2r_3^2+\frac{C^2(r_1^2+r_2^2)r_2^2}{1+C^2r_3^2}\Bigr)\,,\\
2\pi m_2&=&v_2\int_0^{2\pi}\frac{d\sigma}{r_2^2(\sigma)}(1+C^2r_3^2)\,,\\
2\pi m_3&=&v_3\int_0^{2\pi}\frac{d\sigma}{r_3^2(\sigma)}\,.
\end{eqnarray} 
These constraints lead to the undeformed ones when $C=0$ \cite{AFT-N}\,.

\subsection{A deformed S$^3$ subspace}

For simplicity, we shall focus upon a subsector with the geometry R $\times$ deformed S$^3$\,,  
by imposing the condition, 
\begin{eqnarray}
r_2=0\,.
\end{eqnarray}  
Then the effective Lagrangian for $r_3$ and $r_1$ is given by
\begin{eqnarray}
{L}_{\textrm{S}^3}&=&\frac{T_{(C)}}{2}
\Bigl[\dfrac{r_3'^{2}}{1+C^2r_3^2}-r_3^2\omega_3^2- \frac{v_3^2}{r_3^2}
+\dfrac{r_1'^{2}}{1+C^2(1-r_1^2)}-\dfrac{r_1^2\omega_1^2}{1+C^2(1-r_1^2)}\nonumber\\
&&\hspace{1.cm}-\bigl(1+C^2(1-r_1^2)\bigr)\,\frac{v_1^2}{r_1^2}-{\Lambda}(r_3^2+r_1^2-1)\Bigr]\,,
\label{s3NR}
\end{eqnarray} 
where the two constants $v_i$ become
\begin{eqnarray}
v_1=\dfrac{r_1^2\alpha_1'}{1+C^2(1-r_1^2)}\,, \qquad 
v_3={r_3^2\alpha'_3}\,. 
\end{eqnarray} 

\subsubsection*{Virasoro constraints}

The Virasoro constraints are also reduced to 
\begin{eqnarray}
\kappa^2&=& \dfrac{r_3'^{2}}{1+C^2r_3^2}+r_3^2\omega_3^2+ \frac{v_3^2}{r_3^2}+\dfrac{r_1'^{2}}{1+C^2(1-r_1^2)}\nonumber\\
&&\hspace{.cm}+\dfrac{r_1^2\omega_1^2}{1+C^2(1-r_1^2)}+\bigl(1+C^2(1-r_1^2)\bigr)\,\frac{v_1^2}{r_1^2}\,, 
\nonumber \\  
0&=&\omega_3v_3+\omega_1v_1\,. \nonumber 
\end{eqnarray}
The periodicity conditions for $\alpha_i(\sigma)$ become
\begin{eqnarray}
2\pi m_1=v_1\int_0^{2\pi}\frac{d\sigma}{r_1^2(\sigma)}\bigl(1+C^2(1-r_1^2)\bigr)\,,\qquad
2\pi m_3=v_3\int_0^{2\pi}\frac{d\sigma}{r_3^2(\sigma)}\,. \nonumber 
\end{eqnarray}

\subsubsection*{A canonical form}

Finally, let us rewrite the effective Lagrangian (\ref{s3NR}) into a canonical form.

\medskip 

It is convenient to introduce the following new variables:
\begin{eqnarray}
\dfrac{r_3'}{\sqrt{1+C^2r_3^2}}\equiv x_3'\,,\qquad
\dfrac{r_1'}{\sqrt{1+C^2(1-r_1^2)}}\equiv x_1'\,, \nonumber 
\end{eqnarray} 
with the constraint $r_3^2+r_1^2=1$\,. 
By integrating both sides of them, the following relations are obtained, 
\begin{eqnarray}
r_3=\frac{1}{C}\sinh (C x_3)\,,\qquad  r_1=\frac{\sqrt{1+C^2}}{C}\sin (C x_1)\,. \nonumber 
\end{eqnarray} 
Then, in terms of $x_1$ and $x_3$\,, the effective Lagrangian  (\ref{s3NR}) is rewritten as
\begin{eqnarray}
{L}_{\textrm{S}^3}&=&\frac{T_{(C)}}{2}
\Bigl[x_3'^{2}-\frac{\omega_3^2}{C^2}\sinh^2(C x_3)
- \frac{C^2 v_3^2}{\sinh^2 (Cx_3)}+x_1'^{2}-\dfrac{\omega_1^2}{C^2}\tan^2( C x_1)
-\frac{C^2 v_1^2}{\tan^2(C x_1)}\nonumber\\
&&\hspace{1.cm}-\frac{\Lambda}{C^2}\bigl(\sinh^2(C x_3)+(1+C^2)\sin^2( C x_1)-C^2\bigr)\Bigr]\,.
\label{s3NR-canon}
\end{eqnarray} 
Note that the potential terms are expressed with trigonometric and hyperbolic functions. 
This result indicates that the system belongs to the trigonometric class of integrable system. 
By taking the $C\rightarrow 0$ limit, the deformed Lagrangian (\ref{s3NR-canon}) leads to the usual NR system \cite{AFT-N}\,.

\section{Conclusion and discussion}

We have further discussed a $q$-deformation of the AdS$_5\times$S$^5$ superstring. 
In particular, we have argued the nature of the singularity surface 
by adopting a GKP-like string solution as a probe. 
It has been shown that the solution cannot stretch beyond the singularity surface 
and this result may indicate that the holographic relation can be argued only inside 
the singularity surface. According to this observation, we have introduced 
a new coordinate system which describes the spacetime only inside the singularity surface. 
With the new coordinate system, we have studied minimal surfaces and 
derived  a deformed NR system with a rigid string ansatz. 
In particular, the canonical Lagrangian of the resulting system contains 
the interaction potentials of trigonometric and hyperbolic types. 

\medskip 

There are a lot of issues to be studied. 
The most fascinating one is to find out the dual gauge-theory side. 
It would be useful to look for some clues of the corresponding gauge theory 
by employing the new coordinate system. 
The first step is to reveal the geometry around the singularity surface, 
especially with the associated Poincare coordinates.  
We hope that we could report the result in the near future \cite{future}.

\subsection*{Acknowledgments}

We appreciate Gleb Arutyunov, Sergey Frolov, Marc Magro, Takuya Matsumoto,  
Stijn van Tongeren and Benoit Vicedo for useful discussions. 
The work of TK was supported by the Japan Society for the Promotion of Science (JSPS). 
This work is supported in part by JSPS Japan-Hungary Research Cooperative Program.

\section*{Appendix}

\appendix


\section{Trajectories of massless and massive particles}

We consider here trajectories of massless and massive particles 
in order to study the causal structure around the singularity surface both 
in the ABF coordinates (\ref{ads5}) and the new coordinates (\ref{ads5new}). 
As a result, the causal structure around the singularity surface is very similar to  
the boundary of the global AdS space.

\subsection{Trajectories with the ABF metric}

\subsubsection*{A massless particle}

First of all, let us consider a massless particle 
which moves from the origin $\rho=0$ to the singularity surface at $\rho_s=\textrm{arcsinh}(1/C)$\,.

\medskip 

We first consider the trajectory in the coordinate time $t$\,. For simplicity, 
the angle variables are taken to be constant and we drop off the sphere-part contribution. 
Thus the relevant metric is given by   
\begin{eqnarray}
ds^2=R^2(1+C^2)^{\frac{1}{2}}
\Bigl[ -\dfrac{\cosh^2\rho\,dt^2}{1-C^2\sinh^2\rho}
+\dfrac{d\rho^2}{1-C^2\sinh^2\rho}\Bigr]\,.
\end{eqnarray}
Then the trajectory is evaluated from $ds^2=0$\,.

\medskip 

It is convenient to introduce a new variable through $\cos\vartheta=1/\cosh\rho$\,. 
According to this transformation, $\rho\,:0\rightarrow\rho_s$ corresponds to 
$\vartheta\,:0\rightarrow\vartheta_s$\,, where $\vartheta_s=\textrm{arctan}(1/C)$\,. 
Then, by solving the following equation,  
\begin{eqnarray}
ds^2=&R^2(1+C^2)^{\frac{1}{2}}
\Bigl[ \dfrac{-dt^2+d\vartheta^2}{\cos^2\vartheta-C^2\sin^2\vartheta}\Bigr]=0\,,
\end{eqnarray}
the coordinate time from the origin to the singularity is evaluated as follows: 
\begin{eqnarray}
t=\int_0^{\vartheta_s}\!\!d\vartheta=\textrm{arctan}\frac{1}{C}\,.
\label{ABF-t}
\end{eqnarray}
As a result, it does not take infinite time for the massless particle to reach the singularity surface 
in the coordinate time. Note that the well-known result $t=\pi/2$ is reproduced 
in the $C\rightarrow0$ limit.

\medskip

We next consider the trajectory in terms of an affine parameter $t_A$\,. 
The massless condition $p^2=G^{tt}\,p_tp_t + G^{\rho\rho}\,p_{\rho}p_{\rho}=0$ leads to 
the following relation,
\begin{eqnarray}
-\frac{1-C^2\sinh^2\rho}{\cosh^2\rho}\,E^2+\dfrac{1}{1-C^2\sinh^2\rho}
\left(\frac{d\rho}{dt_A}\right)^2=0\,.  
\end{eqnarray}
Here the energy $E$ is defined as $E \equiv -p_t$ and $p_{\rho} \equiv G_{\rho\rho} d\rho/dt_A$.
Then the time from the origin to the boundary is evaluated as 
\begin{eqnarray}
t_A=\int_0^{\rho_s}\frac{d\rho}{E}\dfrac{\cosh\rho}{1-C^2\sinh^2\rho}
=\lim_{\rho\rightarrow\rho_s}\frac{1}{CE}\,\textrm{arctanh}\bigl[C\sinh\rho\bigr] = \infty\,.
\label{ABF-affine}
\end{eqnarray}
Thus it takes infinite affine time to reach the singularity surface at $\rho_s$ as well as 
in the undeformed case. 

\medskip 

In summary, the causal structure around the singularity surface is quite similar to 
the undeformed case. If a reflection boundary condition could be imposed,  
then one may consider the holography only inside the singularity surface. 
To fix the boundary condition precisely, it is necessary to evaluate the energy-momentum tensor 
by explicitly solving the Laplacian on the deformed AdS$_5$\,. 
However, it seems quite intricate and it would need more effort. 
We will leave it as a future problem.

\subsubsection*{A massive particle}

It is turn to consider a trajectory of a massive particle which starts to move from the origin.
The on-shell condition $p^2 = -m^2$ ($m$~:~static mass) leads to the relation, 
\begin{eqnarray}
G^{tt}\,E^2+m^2G_{\rho\rho}\left(\frac{d\rho}{dt_A}\right)^2=-m^2\,.
\end{eqnarray}
Here we have used the relation $p_{\rho} \equiv m G_{\rho\rho}d\rho/dt_A$\,, where $t_A$ is proper time. 
This relation is rewritten as,
\begin{eqnarray}
\left(\frac{d\rho}{dt_A}\right)^2=\frac{\bigl(E^2-m^2-(m^2+C^2E^2)\sinh^2\rho\bigr)(1-C^2\sinh^2\rho)}{m^2\cosh^2\rho}\,.
\label{A8}
\end{eqnarray}
Note that $E \geq m$ because the static mass is $m$. Therefore the r.h.s.\ of (\ref{A8}) is positive at $\rho=0$\,. 
There are two turning points. The one is $\rho=\rho_s$ and the other is 
\[
\rho = \rho_0 \equiv \textrm{arcsinh}\left[\sqrt{\frac{E^2-m^2}{m^2+C^2E^2}}\,\right]\,. 
\]
It is an easy task to show that $\sinh\rho_0 < \sinh\rho_s$ for arbitrary values of $E$\,.
It indicates that, for arbitrary values of $E$\,, the massive particle cannot reach the singularity surface.

\medskip 

By solving the differential equation (\ref{A8})\,, 
one can evaluate the proper time from the origin to the turning point at $\rho=\rho_0$ as follows: 
\begin{eqnarray}
t_A&=&\frac{m}{\sqrt{m^2+C^2E^2}}\int_0^{\rho_0}\!\!\dfrac{d\rho\,\cosh\rho}{\sqrt{(\sinh^2\rho_0-\sinh^2\rho)(1-C^2\sinh^2\rho)}}\nonumber\\
&=&\frac{m}{\sqrt{m^2+C^2E^2}}\, \textrm{K}\Bigl[C^2\sinh^2\rho_0\Bigr]\,.
\label{mABF-affine}
\end{eqnarray}
Notice that (\ref{mABF-affine}) is reduced to $\pi/2$ in the $C\rightarrow0$ limit.

\medskip

The trajectory in the coordinate time is computed from 
the relation 
$dt/dt_A=-G^{tt}E/m$\,. 
Then, by using (\ref{mABF-affine})\,, one can evaluate the coordinate time from the origin to the turning point, 
\begin{eqnarray}
t&=&\frac{E}{\sqrt{m^2+C^2E^2}}\int_0^{\rho_0}\!\!\dfrac{d\rho\,\sqrt{1-C^2\sinh^2\rho}}{\cosh\rho\,\sqrt{\sinh^2\rho_0-\sinh^2\rho}}
\nonumber \\&=&
\frac{(1+C^2)\,E}{\sqrt{m^2+C^2E^2}}\,\Pi\Bigl[-\sinh^2\rho_0, \,C^2\sinh^2\rho_0\Bigr]-\frac{C^2E}{\sqrt{m^2+C^2E^2}}\,\textrm{K}\Bigl[C^2\sinh^2\rho_0\Bigr]\,. \label{mABF-t} 
\end{eqnarray}
Note that (\ref{mABF-t}) becomes $\pi/2$ in the $C\rightarrow0$ limit.

\subsection{The new coordinates}

In the previous subsection, we have studied the trajectories of massless and massive particles 
in the ABF coordinates. Let us here revisit the causal structure around the singularity surface 
in terms of the new coordinates.

\subsubsection*{A massless particle}

The first case is a massless particle 
which moves from the origin $\chi=0$ to the singularity surface at $\chi=\infty$\,.

\medskip

Let us consider the trajectory in the coordinate time $t$\,. 
For simplicity, we drop off the sphere-part as in the previous.  
Then the metric is reduced to 
\begin{eqnarray}
ds^2=&R^2(1+C^2)^{\frac{1}{2}}
\Bigl[ -\cosh^2\chi\,dt^2
+\dfrac{d\chi^2}{1+C^2\cosh^2\chi}\Bigr]=0\,.
\end{eqnarray}
Then the condition $ds^2=0$ leads to the trajectory. 

\medskip

It is convenient to introduce a new variable through  $\cos\theta=1/\cosh\chi$\,. 
According to this transformation, $\chi\,:0\rightarrow\infty$ corresponds to 
$\theta\,:0\rightarrow\pi/2$\,. 
By solving the following equation,  
\begin{eqnarray}
ds^2=&R^2(1+C^2)^{\frac{1}{2}}
\Bigl[ -\dfrac{dt^2}{\cos^2\theta}
+\dfrac{d\theta^2}{\cos^2\theta+C^2}\Bigr]=0\,,
\end{eqnarray}
the coordinate time from the origin to the singularity is evaluated as follows:
\begin{eqnarray}
t=\int_0^{\frac{\pi}{2}}\!\!d\theta\frac {\cos\theta }{\sqrt {\cos^2\theta  +C^2}}=\textrm{arccos}\Bigl[\frac{C}{\sqrt{1+C^2}}\Bigr]\,.
\label{coord-t}
\end{eqnarray}
As a result, it does not take infinite time for the massless particle to reach the singularity surface at $\chi=\infty$
in the coordinate time. Notice that the well-known result $t=\pi/2$ is reproduced 
in the $C\rightarrow0$ limit.

\medskip

Next, we  consider the trajectory in terms of an affine parameter $t_A$\,.
By starting from $p^2=G_{tt}\,p^tp^t + G_{\chi\chi}\,p^{\chi}p^{\chi}=0$\,, 
the following relation is obtained, 
\begin{eqnarray}
-\frac{E^2}{\cosh^2\chi}+\dfrac{1}{1+C^2\cosh^2\chi}\left(\frac{d\chi}{dt_A}\right)^2=0\,,
\end{eqnarray}
with $E \equiv-p_t$\,.
Then the time from the origin to the boundary is evaluated as 
\begin{eqnarray}
t_A=\int_0^{\infty}\frac{d\chi}{E}\dfrac{\cosh\chi}{\sqrt{1+C^2\cosh^2\chi}}=\lim_{\chi\rightarrow\infty}\frac{1}{CE}\,\textrm{arctanh}\Bigl[\frac{C\sinh\chi}{\sqrt{1+C^2\cosh\chi}}\Bigr]\,.
\label{affine-t}
\end{eqnarray}
Thus it takes infinite affine time to reach the singularity surface at $\chi=\infty$ as well as 
in the undeformed case.

\medskip

The above results in the non-zero $C$ case are the standards in the usual AdS, hence it seems likely that 
the singularity surface may be interpreted as the boundary in the new coordinate system.

\subsubsection*{A massive particle}

Next we focus on a trajectory of a massive particle which start to move from the origin.
The on-shell condition $p^2 = -m^2$ ($m$~:~static mass) leads to 
\begin{eqnarray}
G^{tt}\,E^2+m^2G_{\rho\rho}\left(\frac{d\chi}{dt_A}\right)^2=-m^2\,,
\end{eqnarray}
where the energy is defined as $E \equiv -p_t$\,.
This relation is rewritten as 
\begin{eqnarray}
\left(\frac{d\chi}{dt_A}\right)^2=\frac{(E^2-m^2\cosh^2\chi)(1+C^2\cosh^2\chi)}{m^2\cosh^2\chi}\,. 
\label{A18}
\end{eqnarray}
Notice that $E \geq m$ because the static mass is $m$.
Therefore the r.h.s.\ of (\ref{A18}) is positive at $\chi=0$\,. In comparison to the previous, 
there is the unique turning point at 
\[
\chi = \chi_0 \equiv \textrm{arccosh}\left[ \frac{E}{m}\right]\,. 
\]
It indicates that, for arbitrary values of $E$\,, the massive particle cannot reach the singularity surface at $\chi=\infty$\,.

\medskip 

By solving the differential equation (\ref{A18})\,, one can evaluate the proper time from the origin to the turning point like 
\begin{eqnarray}
t_A&=&\int_0^{\chi_0}\!\!\dfrac{d\chi\,\cosh\chi}{\sqrt{(\sinh^2\chi_0-\sinh^2\chi)(1+C^2\cosh^2\chi)}}\nonumber\\
&=&\frac{1}{\sqrt{1+C^2\cosh^2\chi_0}}\,\textrm{K}\Bigl[\frac{C^2\sinh^2\chi_0}{1+C^2\cosh^2\chi_0}\Bigr]\,.
\label{mKame-affine}
\end{eqnarray}
Notice that (\ref{mKame-affine}) is reduced to $\pi/2$ by taking the $C\rightarrow0$ limit.

\medskip

The trajectory in the coordinate time is computed from 
the relation 
$dt/dt_A=-G^{tt}E/m$\,. 
By using the relation in (\ref{mKame-affine})\,, the coordinate time from the origin to the turning point 
is evaluated as
\begin{eqnarray}
t&=&\frac{E}{m}\int_0^{\chi_0}\!\!\dfrac{d\chi}{\cosh\chi\sqrt{(\sinh^2\chi_0-\sinh^2\chi)(1+C^2\cosh^2\chi)}}\nonumber\\&=&
\frac{E}{\sqrt{1+C^2}\,m}\,\Pi\Bigl[-\sinh^2\chi_0,\,-\frac{C^2}{1+C^2}\sinh^2\chi_0\Bigr]\,.
\label{mKame-t}
\end{eqnarray}
Note that (\ref{mKame-t}) is reduced to $\pi/2$ by taking the $C\rightarrow0$ limit.


\section{Giant magnons}

It would be worth revisiting giant magnon solutions \cite{HM,mirror,Kluson,Ahn} 
with the new coordinate system. 
The solution is argued in the $q$-deformed R$\times$S$^3$ subspace 
specified by the condition:
\begin{eqnarray}
\zeta=0\,,\qquad\xi=0\,. 
\label{3d}
\end{eqnarray}
The metric of the deformed S$^3$ part is given by,
\begin{eqnarray}
ds^2_{\textrm{S}^3} &=& R^2(1+C^2)^{\frac{1}{2}}
\Bigl[ \cos^2\gamma\,d\phi^2
+\dfrac{d\gamma^2}{1+C^2\cos^2\gamma}
+\dfrac{\sin^2\gamma\,d\phi_1^2}{1+C^2\cos^2\gamma}
\Bigr]\,.\label{s3}
\end{eqnarray}
Notice that the WZ parts vanish under the condition (\ref{3d})\,.

\subsection{Finite $J$ Giant magnons in R $\times$ deformed S$^3$}

Our argument here basically follows a seminal paper \cite{AFZ} for the undeformed case. 

\medskip 

First of all, we take the following ansatz :
\begin{eqnarray}
t=\kappa\tau\,,\qquad\phi=\phi(\tau,\sigma)\,,\qquad\phi_1=\textrm{const}\,,\qquad\gamma=\gamma(\tau,\sigma)\,.
\label{magnon-s3}
\end{eqnarray}
This ansatz implies
\begin{eqnarray}
E&=&\sqrt{\lambda}\,(1+C^2)^{\frac{1}{2}}\, \kappa\,. \nonumber 
\end{eqnarray}
This relation depends on $C$\,. 

\medskip 

In order to examine motions of $\phi$ and $\gamma$\,, 
let us solve the Virasoro constraints,  
\begin{eqnarray}
\kappa^2&=&\cos^2\gamma\,(\dot{\phi}^2+ \phi'^{2})+\dfrac{\dot{\gamma}^2
+\gamma'^{2}}{1+C^2\cos^2\gamma}\,, \nonumber \\  
0&=&\cos^2\gamma\,\dot{\phi}\,\phi'+\dfrac{\dot{\gamma}\,\gamma'}{1+C^2\cos^2\gamma}\,.
\label{Vira-mag}
\end{eqnarray}
We suppose a solution which satisfies the following boundary conditions:
\begin{eqnarray}
\Delta\phi=\phi(\tau,2\pi)-\phi(\tau,0)=p\,,\qquad 
\Delta\gamma=\gamma(\tau,2\pi)-\gamma(\tau,0)=0\,. \nonumber 
\end{eqnarray}
It is convenient to introduce the following ansatz for  $\phi$ and $\gamma$ :
\begin{eqnarray}
\phi=\tilde{\phi} +\omega\tau\,,\qquad
\tilde{\phi}=\tilde{\phi}(\sigma-v\omega\tau)\,,\qquad
\gamma=\gamma(\sigma-v\omega\tau)\,. \nonumber 
\end{eqnarray}
Then the Virasoro constraints (\ref{Vira-mag}) are rewritten as 
\begin{eqnarray}
\kappa^2&=&\cos^2\gamma\,(\omega-v\omega\partial\tilde{\phi})^2
+\cos^2\gamma\,(\partial\tilde{\phi})^2
+\dfrac{1+v^2\omega^2}{1+C^2\cos^2\gamma}\,(\partial\gamma)^2\,, \nonumber \\  
0&=&\cos^2\gamma\,(\omega-v\omega\partial\tilde{\phi})
\partial\tilde{\phi}-\dfrac{v\omega}{1+C^2\cos^2\gamma}\,(\partial\gamma)^2\,.
\label{Vira-mag2}
\end{eqnarray}
Here $\partial$ means a derivative with respect to $x\equiv \sigma-v\omega\tau$\,.
\begin{eqnarray}
\partial\tilde{\phi}&=&\dfrac{v}{(1-v^2\omega^2)\cos^2\gamma}\,
(\kappa^2-\omega^2\cos^2\gamma)\,, \nonumber \\  
(\partial\gamma)^2&=&\frac{\omega^2(1+C^2\cos^2\gamma)}{(1-v^2\omega^2)^2\cos^2\gamma}
\Bigl(\frac{\kappa^2}{\omega^2}-\cos^2\gamma \Bigr)(\cos^2\gamma-v^2\kappa^2 )\,. \nonumber 
\end{eqnarray}
It is helpful to introduce a radial coordinate $z$ as $z=\sin\gamma$\,.

\medskip 

Then above equations can be rewritten as
\begin{eqnarray}
\partial\tilde{\phi}&=&\dfrac{v\omega^2}{(1-v^2\omega^2)}\frac{z^2-z^2_{\rm min}}{1-z^2}\,, \nonumber \\  
(\partial z)^2&=&\frac{\omega^2(1+C^2-C^2z^2)}{(1-v^2\omega^2)^2}
(z^2-z_{\rm min}^2)(z_{\rm max}^2-z^2 )\,, \nonumber 
\end{eqnarray}
where we have introduced the following quantities, 
\begin{eqnarray}
z_{\rm min}^2=1-\frac{\kappa^2}{\omega^2}\,,\qquad z_{\rm max}^2=1-v^2\kappa^2\,.
\end{eqnarray}
Here $v<1/\kappa$ is assumed.

\medskip

Next let us examine the conserved charge $J$ :
\begin{eqnarray}
J&=&\sqrt{\lambda}\,(1+C^2)^{\frac{1}{2}}\int^{2\pi}_{0}\frac{d\sigma}{2\pi}\,(1-z^2)\, \dot{\phi}\,.
\end{eqnarray}
Note that, by using $d\sigma=dz/z'$\,, the following relation is obtained, 
\begin{eqnarray}
\!\!\!\!2\pi=\int^{2\pi}_{0}\!\!d\sigma=2\int^{z_{\rm max}}_{z_{\rm min}}\frac{dz}{|z'|}=\int^{z_{\rm max}}_{z_{\rm min}}\!\!
\frac{2(1-v^2\omega^2)}{\omega\sqrt{1+C^2-C^2z^2}}
\frac{dz}{\sqrt{(z^2-z_{\rm min}^2)(z_{\rm max}^2-z^2 )}}\,.
\end{eqnarray}
Then $J$ can be expressed as
\begin{eqnarray}
J=\frac{\sqrt{\lambda}}{\pi}\,(1+C^2)^{\frac{1}{2}}\int^{z_{\rm max}}_{z_{\rm min}}\frac{dz}{\sqrt{1+C^2-C^2z^2}}
\frac{z_{\rm max}^2-z^2}{\sqrt{(z^2-z_{\rm min}^2)(z_{\rm max}^2-z^2 )}}\,.
\end{eqnarray}
The magnon momentum $p$ is also given by
\begin{eqnarray}
p&=&\int^{2\pi}_{0}\!\!d\sigma\,\phi^{'}=2\int^{z_{\rm max}}_{z_{\rm min}}\frac{dz}{|z'|}\,\partial\tilde{\phi}\nonumber\\
&=&2v\omega\int^{z_{\rm max}}_{z_{\rm min}}\frac{dz}{\sqrt{1+C^2-C^2z^2}(1-z^2)}
\frac{z^2- z_{\rm min}^2}{\sqrt{(z^2-z_{\rm min}^2)(z_{\rm max}^2-z^2 )}}\,.
\end{eqnarray}
In principle, it would be possible to derive the dispersion relation of the magnon, 
but we will not do here. It is beyond our scope and we are interested in a feature 
of the new coordinate system. Hence, for our purpose, it would be 
rather useful to consider the infinite $J$ limit and 
focus upon a simper expression of the dispersion relation.

\subsection{Infinite $J$ giant magnon}

Let us consider the infinite $J$ giant magnon.
It corresponds to the case with $z_{\rm min}=0$\,, i.e. 
\begin{eqnarray}
\kappa=\omega\,,\qquad z_{\rm max}^2=1-v^2\omega^2\,. \nonumber 
\end{eqnarray}
The magnon momentum $p$ becomes
\begin{eqnarray}
p&=&\int^{2\pi}_{0}\!\!d\sigma\,\phi'=2\int^{z_{\rm max}}_{0}\frac{dz}{|z'|}\,\partial\tilde{\phi}\nonumber\\
&=&2v\omega\int^{z_{\rm max}}_{0}\frac{dz}{\sqrt{1+C^2-C^2z^2}}\frac{z}{1-z^2}
\frac{1}{\sqrt{z_{\rm max}^2-z^2 }} \nonumber\\
&=&i\log\Bigl[1- \frac{2z_{\rm max}^2+2i(1+C^2)^{\frac{1}{2}}
\sqrt{z_{\rm max}^2-z_{\rm max}^4}}{1+C^2(1-z_{\rm max}^2)}\Bigr]\,. \nonumber 
\end{eqnarray}
This leads to the following relation, 
\begin{eqnarray}
\sin^2\frac{p}{2}=\frac{1-v^2\omega^2}{1+C^2v^2\omega^2}\,.
\label{momentum}
\end{eqnarray}
The periodicity is expressed as 
\begin{eqnarray}
2\pi=\int^{z_{\rm max}}_{0}\!\!
\frac{2(1-v^2\omega^2)}{\omega\sqrt{1+C^2-C^2z^2}}
\frac{dz}{z\sqrt{z_{\rm max}^2-z^2 }}\,. \nonumber 
\end{eqnarray}

\medskip 

Then the energy $E$ is given by 
\begin{eqnarray}
E=\frac{\sqrt{\lambda}}{\pi}\,(1+C^2)^{\frac{1}{2}}\int^{z_{\rm max}}_{0}\!\!
\frac{(1-v^2\omega^2)}{\sqrt{1+C^2-C^2z^2}}
\frac{dz}{z\sqrt{z_{\rm max}^2-z^2 }}\,, \nonumber 
\end{eqnarray}
and $J$ becomes
\begin{eqnarray}
J&=&\frac{\sqrt{\lambda}}{\pi}\,(1+C^2)^{\frac{1}{2}}\int^{z_{\rm max}}_{0}\frac{dz}{\sqrt{1+C^2-C^2z^2}}
\frac{\sqrt{z_{\rm max}^2-z^2}}{z}\,. \nonumber 
\end{eqnarray}
By using (\ref{momentum})\,, the giant magnon dispersion relation can be expressed as
\begin{eqnarray}
E-J=\frac{\sqrt{\lambda}}{\pi}\frac{\sqrt{1+C^2}}{C}\,\textrm{arctanh}\frac{C\, |\sin\frac{p}{2}|}{\sqrt{1+C^2\sin^2\frac{p}{2}}}\,.
\label{dispersion}
\end{eqnarray}
This result exactly agrees with the  dispersion relation derived in \cite{mirror,Kluson}\,. 

\medskip 

By converting $C$ into $\nu$ through 
\begin{eqnarray}
\nu 
=\frac{C}{\sqrt{1+C^2}}\,, \nonumber 
\end{eqnarray}
the relation (\ref{dispersion}) can be expressed as
\begin{eqnarray}
E-J=\frac{\sqrt{\lambda}}{\pi\nu}\, \textrm{arcsinh}\frac{\nu\, |\sin\frac{p}{2}|}{\sqrt{1-\nu^2}}\,. \nonumber 
\end{eqnarray}

\section{Another deformed NR system}

We derive here another deformation of the NR system 
from the deformed AdS$_5$ (For the metric, see (\ref{ads5new}))\,. 
The analysis is quite parallel to the deformed S$^5$ case. 

\subsection{A rigid string ansatz} 

Let us first rewrite the metric of the deformed AdS$_5$ in (\ref{ads5new}) with the following coordinates:
\begin{eqnarray}
s_0=\cosh\chi\,,\qquad s_1=\sinh\chi\cos\zeta\,,\qquad s_2=\sinh\chi\sin\zeta\,.
\end{eqnarray} 
Then the resulting metric is given by 
\begin{eqnarray}
ds^2_{\textrm{AdS}_5} &=& R^2(1+C^2)^{\frac{1}{2}}
\Bigl[ -\dfrac{ds_0^2}{1+C^2s_0^2}-s_0^2\,dt^2
+\dfrac{ds_1^2}{1+C^2s_0^2}+\dfrac{(1+C^2s_0^2)s_1^2\,d\psi_1^2}{(1+C^2s_0^2)^2+C^2(s_1^2+s_2^2)s_2^2}
\nonumber\\ 
&&\hspace{.cm}+\dfrac{ds_2^2+s_2^2\,d\psi_2^2}{1+C^2s_0^2}
-\dfrac{C^2s_2^2(s_1ds_2-s_2ds_1)^2}{(1+C^2s_0^2)[(1+C^2s_0^2)^2+C^2(s_1^2+s_2^2)s_2^2]}
\Bigr]\,.\label{AdS5}
\end{eqnarray}
The WZ part can also be rewritten as
\begin{eqnarray}
\mathcal{L}_{\textrm{AdS}}^{\rm WZ} &=& \frac{T_{(C)}}{2}\,
\epsilon^{\mu\nu}\dfrac{2C s_1s_2\,\partial_\mu\psi_1
(s_1\partial_\nu s_2-s_2\partial_\nu s_1)}{(1+C^2s_0^2)^2+C^2(s_1^2+s_2^2)s_2^2}\,.
\label{WZ_ads}
\end{eqnarray}

\medskip

Suppose a rigid string ansatz : 
\begin{eqnarray}
&&t=\nu_0\,\tau+\beta_0(\sigma)\,,\quad\psi_1=\nu_1\,\tau+\beta_1(\sigma)\,,
\quad\psi_2=\nu_2\,\tau+\beta_2(\sigma)\,, \quad 
\nonumber\\ &&
\chi=\chi(\sigma)\,,
\quad\zeta=\zeta(\sigma)\,.
\label{rigid_ads}
\end{eqnarray}
Here $\nu_i~(i=0,1,2)$ are real constant parameters. 
According to this ansatz, the periodic boundary conditions are translated 
into the following form, 
\begin{eqnarray}
&&s_a(\sigma+2\pi)=s_a(\sigma)\qquad(a=0,1,2)\,,\\
&&\beta_0(\sigma+2\pi)=\beta_0(\sigma)\,, \label{2nd} \\
&&\beta_{a'}(\sigma+2\pi)=\beta_{a'}(\sigma)+2\pi k_{a'}\qquad 
(k_{a'} \in \mathbb{Z}\,, \quad a'=1,2)\,.
\end{eqnarray}
Note that the second condition (\ref{2nd}) follows from the requirement that 
$t$ should be single valued for $\sigma$\,.

\subsection{Reduction to a NR system}

With the ansatz (\ref{rigid_ads})\,, the Lagrangian results in a 1-d dynamical system, 
\begin{eqnarray}
\mathcal{L}_{\textrm{AdS}_5}&=&-\frac{T_{(C)}}{2}
\Bigl[ 
-\dfrac{s_0'^{2}}{1+C^2s_0^2}-s_0^2(-\nu_0^2+\beta_0'^{2})+\dfrac{s_1'^{2}}{1+C^2s_0^2} 
+\dfrac{(1+C^2s_0^2)s_1^2(-\nu_1^2+\beta_1'^{2})}{(1+C^2s_0^2)^2+C^2(s_1^2+s_2^2)s_2^2}\nonumber\\ &&\hspace{1cm}
+\dfrac{s_2'^{2}+s_2^2(-\nu_2^2+\beta_2'^{2})}{1+C^2s_0^2}
-\dfrac{C^2s_2^2(s_1s'_2-s_2s'_1)^2}{(1+C^2s_0^2)[(1+C^2s_0^2)^2+C^2(s_1^2+s_2^2)s_2^2]}\nonumber\\ &&\hspace{1cm}
-\dfrac{2C\nu_1 s_1s_2(s_1s'_2-s_2s'_1)}{(1+C^2s_0^2)^2+C^2(s_1^2+s_2^2)s_2^2}-\widetilde{\Lambda}(\eta^{ab}s_as_b+1)\Bigr]\,,
\label{ads5_rigid}
\end{eqnarray} 
where $\widetilde{\Lambda}$ is  a Lagrange multiplier and $\eta^{ab}=$ diag$(-1,+1,+1)$\,. 

\medskip 

The equations of motion for $\beta_a(\sigma)$ lead to 
\begin{eqnarray}
u_0\equiv s_0^2\,\beta'_0\,,\quad u_1\equiv\dfrac{(1+C^2s_0^2)\,s_1^2\,\beta'_1}{(1+C^2s_0^2)^2+C^2(s_1^2+s_2^2)\,s_2^2}\,, 
\quad u_2\equiv\dfrac{s_2^2\,\beta'_2}{1+C^2s_0^2}\,,\quad u_a=\textrm{const}. \nonumber 
\end{eqnarray} 
Here $u_a$ are three integration constants. 
Note that $u_a$ depend on $C$\,.
When $C=0$\,, the undeformed results are reproduced \cite{AFT-N}\,.

\medskip

By deleting $\beta_a$ from (\ref{ads5_rigid})\,, the effective Lagrangian for $s_a$ is obtained as
\begin{eqnarray}
{L}_{\textrm{AdS}_5}&=&\frac{T_{(C)}}{2}
\Bigl[-\dfrac{s_0'^{2}}{1+C^2s_0^2}+s_0^2\nu_0^2+ \frac{u_0^2}{s_0^2}+\dfrac{s_1'^{2}}{1+C^2s_0^2}
-\dfrac{(1+C^2s_0^2)s_1^2\nu_1^2}{(1+C^2s_0^2)^2+C^2(s_1^2+s_2^2)s_2^2}\nonumber\\
&&\hspace{.cm}-\Bigl(1+C^2s_0^2+\frac{C^2(s_1^2+s_2^2)s_2^2}{1+C^2s_0^2}\Bigr)\,\frac{u_1^2}{s_1^2}
+\dfrac{s_2'^{2}-s_2^2\nu_2^2}{1+C^2s_0^2}-(1+C^2s_0^2)\frac{u_2^2}{s_2^2}\nonumber\\ 
&&\hspace{.cm}-\dfrac{C^2s_2^2(s_1s'_2-s_2s'_1)^2}{(1+C^2s_0^2)[(1+C^2s_0^2)^2+C^2(s_1^2+s_2^2)s_2^2]}
-\dfrac{2C\nu_1 s_1s_2(s_1s'_2-s_2s'_1)}{(1+C^2s_0^2)^2+C^2(s_1^2+s_2^2)s_2^2}\nonumber\\
&&\hspace{.cm}-\widetilde{\Lambda}(\eta^{ab}s_as_b+1)\Bigr]\,.
\label{ads5NR}
\end{eqnarray} 
Here we have changed the sign of (\ref{ads5_rigid}), except for the terms which include $ u_a^2/ s_a^2$ 
so as to reproduce the same equations of motion for $s_a$\,. 
As a result, the Lagrangian (\ref{ads5NR}) can be regarded as a deformed 
NR system. When $C=0$\,, this system reduces to the usual NR system \cite{AFT-N}\,.

\subsubsection*{Virasoro constraints}

Finally, we shall comment on the Virasoro constraints. These become
\begin{eqnarray}
\dfrac{s_0'^{2}}{1+C^2s_0^2}+s_0^2\nu_0^2+\frac{u_0^2}{s_0^2}&=& \dfrac{s_1'^{2}}{1+C^2s_0^2}+\dfrac{(1+C^2s_0^2)s_1^2\nu_1^2}{(1+C^2s_0^2)^2+C^2(s_1^2+s_2^2)s_2^2}\nonumber\\
&&\hspace{-2.5cm}+\Bigl(1+C^2s_0^2+\frac{C^2(s_1^2+s_2^2)s_2^2}{1+C^2s_0^2}\Bigr)\,\frac{u_1^2}{s_1^2}
+\dfrac{s_2'^{2}+s_2^2\nu_2^2}{1+C^2s_0^2}+(1+C^2s_0^2)\frac{u_2^2}{s_2^2}\,,\label{V1}\\  
0&=&\eta^{ab}\nu_au_b\,.\label{V2}
\end{eqnarray}
From the periodicity conditions for $\beta_a(\sigma)$\,, one can obtain the following relations:
\begin{eqnarray}
2\pi k_1&=&u_1\int_0^{2\pi}\frac{d\sigma}{s_1^2(\sigma)}\Bigl(1+C^2s_0^2
+\frac{C^2(s_1^2+s_2^2)s_2^2}{1+C^2s_0^2}\Bigr)\,, \nonumber \\
2\pi k_2&=&u_2\int_0^{2\pi}\frac{d\sigma}{s_2^2(\sigma)}(1+C^2s_0^2)\,, \nonumber \\
0&=&u_0\int_0^{2\pi}\frac{d\sigma}{s_0^2(\sigma)}\,. \nonumber 
\end{eqnarray} 
It is necessary to set $u_0=0$ so that $t$ should be single valued. 
When $C=0$\,, these constraints lead to the ones for the undeformed case \cite{AFT-N}\,.

\subsection{A deformed AdS$_3$ subspace}

For simplicity, we will focus upon a deformed AdS$_3\times$S$^1$
subsector of the $q$-deformed AdS$_5\times$S$^5$ by imposing the following condition, 
\begin{eqnarray}
s_2=0\,.
\end{eqnarray}  
Then the effective Lagrangian for $s_0$ and $s_1$ is given by
\begin{eqnarray}
{L}_{\textrm{AdS}_3}&=&\frac{T_{(C)}}{2}
\Bigl[-\dfrac{s_0'^{2}}{1+C^2s_0^2}+s_0^2\nu_0^2+ \frac{u_0^2}{s_0^2}
+\dfrac{s_1'^{2}}{1+C^2(1+s_1^2)}-\dfrac{s_1^2\nu_1^2}{1+C^2(1+s_1^2)}\nonumber\\
&&\hspace{1.cm}-\bigl(1+C^2(1+s_1^2)\bigr)\,\frac{u_1^2}{s_1^2}-\widetilde{\Lambda}(-s_0^2+s_1^2+1)\Bigr]\,,
\label{ads3NR}
\end{eqnarray} 
where the remaining two constants $u_a$ are 
\begin{eqnarray}
u_0={s_0^2\beta'_0}=0\,,\qquad u_1=\dfrac{s_1^2\beta'_1}{1+C^2(1+s_1^2)}=\textrm{const}\,.
\end{eqnarray}

\subsubsection*{Virasoro constraints}

It is turn to see the Virasoro constraints for the 
subsector. 
By adding an angle $\phi$\,, which describes S$^1$\,,
\begin{eqnarray}
\phi= \omega_3\tau+m_3\sigma\qquad 
 \qquad (m_3\in \mathbb{Z})\,, 
\end{eqnarray}
the Virasoro constraints are recast into the form, 
\begin{eqnarray}
\dfrac{s_0'^{2}}{1+C^2s_0^2}+s_0^2\nu_0^2
&=& \dfrac{s_1'^{2}}{1+C^2(1+s_1^2)}+\dfrac{s_1^2\nu_1^2}{1+C^2(1+s_1^2)}
+\bigl(1+C^2(1+s_1^2)\bigr)\,\frac{u_1^2}{s_1^2}+\omega_3^2+m_3^2
\,,\nonumber\\  
0&=&\nu_1u_1+\omega_3m_3\,.
\end{eqnarray}
The periodicity condition for $\beta_1(\sigma)$ is given by
\begin{eqnarray}
2\pi k_1&=&u_1\int_0^{2\pi}\frac{d\sigma}{s_1^2(\sigma)}\bigl(1+C^2(1+s_1^2)\bigr)\,.
\end{eqnarray}

\subsubsection*{A canonical form}

Finally, let us rewrite the effective Lagrangian (\ref{ads3NR}) into the canonical form.

\medskip 

It is convenient to introduce the following new variables:
\begin{eqnarray}
\dfrac{s_0'}{\sqrt{1+C^2s_0^2}}\equiv y_0'\,,\qquad
\dfrac{s_1'}{\sqrt{1+C^2(1+s_1^2)}}\equiv y_1'\,, \nonumber 
\end{eqnarray} 
with the constraint $s_0^2-s_1^2=1$\,. 
Integrating them leads to the following relations 
\begin{eqnarray}
s_0=\frac{1}{C}\sinh( C y_0)\,,\qquad
s_1=\frac{\sqrt{1+C^2}}{C}\sinh( C y_1)\,. \label{A25}
\end{eqnarray} 
Then, after putting (\ref{A25}) into (\ref{ads3NR}), 
the resulting Lagrangian is given by 
\begin{eqnarray}
{L}_{\textrm{AdS}^3}&=&\frac{T_{(C)}}{2}
\Bigl[-y_0'^{2}+\frac{\nu_0^2}{C^2}\sinh^2(C y_0) 
+y_1'^{2}-\dfrac{\nu_1^2}{C^2}\tanh^2 (C y_1)-\frac{C^2 u_1^2}{\tanh^2 (Cy_1)}\nonumber\\
&&\hspace{1.cm}-\frac{\widetilde{\Lambda}}{C^2}\bigl(-\sinh^2(C y_0)+(1+C^2)\sinh^2 (C y_1)-C^2\bigr)\Bigr]\,.
\label{ads3NR-canon}
\end{eqnarray} 
The potential terms are of hyperbolic type. 
Taking the $C\rightarrow0$ limit, the Lagrangian (\ref{ads3NR-canon}) 
is reduced to the usual NR system \cite{AFT-N}.

\end{document}